\documentclass[aps,prb,twocolumn,amsmath]{revtex4}
\usepackage[dvips]{color}
\usepackage[normalem]{ulem}

\definecolor{g-blue}{rgb}{0.83,0.95,1}
\definecolor{g-yellow}{rgb}{1,1,0.7}
\definecolor{g-green}{rgb}{0.9,1,0.9}
\definecolor{green}{rgb}{0,0.6,0}
\definecolor{cyan}{rgb}{0,0.7,0.7}
\definecolor{black}{rgb}{0,0,0}
\definecolor{grey}{rgb}{0.4 ,0.4 ,0.4 }

\usepackage{bm}
\usepackage{amssymb}
\usepackage{amsfonts}
\usepackage{amsmath}
 \usepackage{graphicx}
  \usepackage{amsmath,bm,epsfig}
\def \ed {\end{document}}
\def\Fbox#1{\vskip1ex\hbox to 8.5cm{\hfil\fboxsep0.3cm\fbox{%
  \parbox{8.0cm}{#1}}\hfil}\vskip1ex\noindent}  

\newcommand{\eq}[1]{(\ref{#1})}
\newcommand{\Eq}[1]{Eq.\,(\ref{#1})}
\newcommand{\Eqs}[1]{Eqs.\,(\ref{#1})}
\newcommand{\Fig}[1]{Fig.\,\ref{#1}}
\newcommand{\Figs}[1]{Figs.\,\ref{#1}}
\newcommand{\Sec}[1]{Sec.\,\ref{#1}}
\newcommand{\Ref}[1]{Ref.\,\cite{#1}}
\newcommand{\Refs}[1]{Refs.\,\cite{#1}}

\def\be{\begin{equation}}\def\ee{\end{equation}}
\def\bea{\begin{eqnarray}}\def\eea{\end{eqnarray}}
\def\bse{\begin{subequations}}\def\ese{\end{subequations}}
\newcommand{\BE}[1] {\begin{equation}\label{#1}}
\newcommand{\BEA}[1]{\begin{eqnarray}\label{#1}}
\newcommand{\BSE}[1]{\begin{subequations}\label{#1}}

\let \nn  \nonumber  \newcommand{\br}{\\ \nn}

\let \= \equiv  \let\*\cdot \let\~\widetilde \let\^\widehat \let\-\overline
\let\p\partial

  \def\1{\bm1} 

\def\<{\left\langle}    \def\>{\right\rangle}
\def\({\left(}          \def\){\right)}
 \def \[ {\left [} \def \] {\right ]}

\renewcommand{\a}{\alpha}\renewcommand{\b}{\beta}\newcommand{\g}{\gamma}
\renewcommand{\d}{\delta}
\newcommand{\ve}{\varepsilon}
\renewcommand{\o}{\omega} \renewcommand{\O}{\Omega}

\newcommand{\B}[1]{{\bm{#1}}}
\newcommand{\C}[1]{{\mathcal{#1}}}    

\renewcommand{\sb}[1]{_{\text {#1}}}  
\renewcommand{\sp}[1]{^{\text {#1}}}  
\newcommand{\Sp}[1]{^{^{\text {#1}}}} 
\def\Sb#1{_{\scriptscriptstyle\rm{#1}}}
\def\He4 {$^4$He~}\def\He3 {$^3$He~}

\begin{document}

\title{ Local and non-local energy spectra of superfluid  $^3$He turbulence }
\author{L. Biferale, D. Khomenko, V. L'vov, A. Pomyalov, I. Procaccia and G. Sahoo}

\begin{abstract}
Below the phase transition temperature $T\sb c\simeq 10^{-3}\,$K $^3$He-B has a mixture of normal and superfluid components. Turbulence in this material is carried predominantly by the superfluid component. We explore  the statistical properties of this quantum turbulence, stressing the differences from the better known classical counterpart. To this aim we study the time-honored Hall-Vinen-Bekarevich-Khalatnikov coarse-grained equations of superfluid turbulence. We combine pseudo-spectral direct numerical simulations with analytic considerations based on an integral closure for the energy flux. We avoid the assumption of locality of the energy transfer  which was used previously in both analytic and numerical studies of the superfluid $^3$He-B  turbulence.
For $T<0.37 \, T\sb c$, with relatively weak mutual friction, we confirm the previously found ``subcritical" energy spectrum $E(k)$,  {given by a superposition of two power laws that can be approximated as  $E(k)\propto k^{-x}$ with an apparent scaling exponent $\frac 53 <x(k)< 3$}. For  $T>0.37 \, T\sb c$ and with strong mutual friction, we observed numerically and confirmed analytically the scale-invariant spectrum $E(k)\propto k^{-x}$ with  a ($k$-independent) exponent $x > 3$ that gradually increases with the temperature and reaches a value $x\sim 9$  for $T\approx 0.72\, T\sb c$. In the near-critical regimes we discover  a strong  enhancement of intermittency which exceeds by an order of magnitude the corresponding level in classical hydrodynamic turbulence.
\end{abstract}

\maketitle

\section*{Introduction}

Helium below the phase transition temperatures $T_\lambda \simeq 2.1\,$K in $^4$He and $T\sb c\simeq 10^{-3}\,$K in $^3$He can be described as consisting of two coupled, interpenetrating
fluids. One fluid is inviscid with quantized vorticity, and the second is viscous with a continuous vorticity. Consequently, superfluid turbulence is even more complex than turbulence in classical fluids. Moreover, the present knowledge of many aspects of superfluid turbulence is still not fully developed despite the many
decades since the discovery of superfluidity, see, e.g. \Refs{Rev1,Rev2,Rev3,Rev4}. The subject offers
many opportunities for new approaches and new discoveries. 

From the experimental point of view the study of the statistical properties of superfluid turbulence is still difficult, even with the use of state-of-the-art technologies.
The very low values of $T_\lambda$ and  $T\sb c$ limit severely any visual access, and in addition pose problems for adequate sensors\,\cite{Rev1,Rev2,Rev3,Rev4}. Nevertheless new  experiments are emerging, requiring parallel
theoretical efforts. Theoretical progress requires developing direct numerical simulations (DNS) which
presently are the only way to reach a complete description of the evolution of the normal and superfluid  velocity components. Such data offer access to the statistical properties of superfluid turbulence. In the present paper we study the physics of superfluid $^3$He turbulence, using the fact that it is simpler problem than turbulence in $^4$He, due to very high viscosity of the normal component, which may be considered laminar.

The energy spectra $E(k)$ in space-homogeneous, steady and isotropic turbulence in superfluid $^3$He were studied analytically within the algebraic approximation for the energy flux in \Ref{LNV} (see also Eq.\,\eqref{alg} below). Numerically the issue was studied using the Sabra-shell model in \Ref{He3a}. The two papers ~\cite{LNR,He3a} considered the large-scale velocity fluctuations  with $k< \pi /\ell$, where $\ell$ is the mean distance between quantized vortex lines.  It was shown  that the mutual friction between normal and superfluid components suppresses $ E(k)$  with respect of the Kolmogorov-1941 (K41) prediction\,\cite{Fri}:
\begin{equation}\label{K41}
 E\Sp{K41} (k)= C\Sb K \ve_0^{2/3} k^{-5/3}\ .
\end{equation}
Here $\ve_0$ is the energy flux over scales, equal in this case to the rate of energy input into the system at $k=k_0$: $\ve_0=\ve(k_0)$; $C\Sb K \sim 1$ is the dimensionless Kolmogorov constant.

The isotropic, steady-state energy balance equation in a one-fluid approach to \He3 turbulence was analyzed by Lvov, Nazarenko and Volovik (LNV) in \Ref{LNV}:
\begin{subequations} \label{bal}
\begin{equation}\label{balA}
\frac{d \,\ve (k)} {d k}+ \Omega \,  E(k)=0\,, \quad \Omega = \alpha(T)\, \Omega\Sb T\ .
\end{equation}
Here $\alpha(T)$ is the temperature dependent dimensionless mutual friction parameter and $\Omega\Sb T$ is the root mean square (rms) turbulent vorticity. The wavenumber-dependent energy flux over scales, $\ve(k)$, was approximated in \Ref{LNV} using K41-type dimensional reasoning, similar to \Eq{K41}:
\begin{equation}\label{alg}
\ve(k)= \big [  E(k)\big / C \Sb{K}\big ]^{3/2}\, k^{5/2}= \frac 83 \big [  E(k) \big ]^{3/2}\, k^{5/2}\,,
\end{equation}
as suggested  by Kovasznay\,\cite{Kov}.
\end{subequations}

The ordinary differential \Eq{bal} has an analytical solution\,\cite{LNV}:
\begin{subequations}\label{Spectra}\begin{eqnarray}\label{LNV-a}
  E (k)&=&  E\Sp{K41}(k) \Big [ 1- \Omega^\dag  +\Omega^\dag  \Big( \frac {k_0}k\Big )^{2/3}\Big ]^2,\ \mbox{where}~~~~~~~~\\ \label{LNV-b}
\Omega^\dag &=&\frac{\O}{\Omega\sb{cr}} \,, \quad \Omega\sb {cr}=\frac 54\, \sqrt{k_0^3 \, E_0}\,, \quad   E_0\=  E(k_0)\ .
 \eea
 For $\Omega^\dag <1$ \Eq{LNV-a} introduces a new crossover length-scale
\begin{equation}\label{LNV-c}
k_\times=k_0 \big [\Omega^\dag\big / (1-\Omega^\dag)   \big ]^{3/2}\,,
\end{equation}
that breaks the scaling invariance, predicting for $\Omega<\Omega\sb {cr}$ a superposition of two scaling laws:\\
  -- For small $k\ll k_\times$,  the LNV spectrum\,\eq{LNV-a}  takes a ``critical" form
\begin{equation}\label{LNV-cr}
E \sb{cr}(k)=  E_0 \big(k_0/k\big )^{3}\ .
\end{equation}\\
-- For large enough $k\gg k_\times$, the K41 spectrum\,\eqref{K41}  is recovered, but with the energy flux $\ve_\infty< \ve_0$. The difference $\ve_0-\ve_\infty$ is dissipated by the mutual friction.  For $k\sim k_\times$,
    the energy spectrum can be roughly approximated as  $E(k)\propto k^{-x}$ with an apparent scaling exponent $\frac53 < x(k) < 3$.

The crossover wavenumber $k_\times$ increases with $\alpha(T)$ and for some critical value of $\alpha\sb {cr}\sim 1$ it diverges. Then the critical LNV-spectrum\,\eqref{LNV-cr} occupies the entire available interval $k_0< k < \pi/\ell$.

For $\alpha(T) > \alpha\sb {cr}$, the spectrum\,\eqref{LNV-a} becomes ``supercritical" and terminates at some final $k_*$ that depends on $\alpha(T)$:
\begin{equation}\label{LNV-sup}
E\sb s(k)\propto k^{-3}\big[k_*^{2/3}- k^{2/3}\big]^2\ .
\end{equation}\end{subequations}
All types of the LNV spectra  (subcritical, critical and supercritical) where observed in Sabra-shell model simulations (see \Refs{He3a} and \cite{lb} for a general review on shell models).
However, the analytical LNV model\,\cite{LNV} is based on an uncontrolled algebraic approximation for the energy flux\,\eqref{alg};  the shell-model of turbulence, used in \Ref{He3a},  is also an uncontrolled simplification of the basic equations of motion for the superfluid velocity field.  Therefore, the problem of turbulent energy spectra in superfluid $^3$He requires further investigation.

 In this paper we report results of a first (to the best of our knowledge) DNS study of the statistical properties of a space-homogeneous, steady and isotropic turbulence in superfluid $^3$He. We provide results on the turbulent energy spectra, the  velocity and vorticity structure functions at different temperatures $0< T < 0.7 T\sb{cr}$, the energy balance and intermittency effects. To these aims we use the gradually-damped version of  the Hall-Vinen\,\cite{HV}-Bekarevich-Khalatnikov\cite{BK} (HVBK) coarse-grained two-fluid model  Eq.\,\eqref{NSE} as suggested in \Ref{PRB}. We expect this model to describe properly the turbulent velocity fluctuations in superfluid $^4$He and $^3$He as long as the their scales exceed the mean intervortex distance $\ell$.

  The paper is organized as follows: \\~\\
  \textbullet~Section\,\ref{s:He3}  is devoted to an analytical description of the statistical properties of the steady, homogeneous, isotropic, incompressible turbulence of superfluid $^3$He. This should serve as a basis for further studies of superfluid turbulence in more complicated or/and realistic cases: anisotropic turbulence, transient regimes, two-fluid turbulence of counterflowing, thermally driven, superfluid  $^4$He turbulence, etc.
  \begin{description}
  \item
  In \Sec{ss:HVBK} we present the gradually damped HVBK \Eqs{NSE};
   \item
   In \Sec{ss:defs} we introduce the required statistical objects.
  \item In \Sec{ss:integral}  we adapt the  integral closure\cite{LNR} to obtain the energy spectrum when the energy transfer over scales is not local.

\item In \Sec{sss:struct} we analyze the relations between the structure functions of the velocity and vorticity fields with the sub- and super-critical energy spectra $ E(k)$. These are required for the analysis of the DNS data.

\end{description}

  \textbullet~Section\,\ref{s:DNS} presents the DNS results for the statistics of superfluid turbulence in $^3$He, together with a comparison with the theoretical expectations.

  \begin{description}
  \item In \Sec{ss:procedure} we shortly describe the details of the numerical procedure;
 \item In \Sec{ss:energy} we present the DNS results for the energy spectra obtained for different values of mutual friction frequency $\Omega$ in the subcritical, critical and supercritical regimes. We demonstrate their quantitative agreement with the corresponding theoretical predictions, given by \Eqs{LNV-a}, \eqref{LNV-cr} and \eqref{28};
  \item In \Sec{ss:str} we  report a significant enhancement of intermittency in near-critical  regimes of
       superfluid $^3$He turbulence, revealed by analysing the second- and fourth-order structure functions of the velocity and vorticity differences;

  \item In \Sec{ss:balance} we analyze the energy balance in the entire region of $k$, shedding light on the origin of the subcritical, critical and supercritical regimes of the energy spectra;
  \item In \Sec{ss:evolution} we present and analyze the DNS results for the energy and enstropy time evolution, showing how the large and small scale turbulent fluctuations are correlated (or uncorrelated) in different regimes;

    \item Section  \ref{ss:temp} clarifies the relation between the mutual friction frequency $\O$ and the temperature $T$ in possible experiments.
 \end{description}

 \textbullet~Section\,\ref{s:sum} summarizes our findings. For the convenience of the reader we present here
the main results:
 \begin{description}
 \item The  numerical subcritical energy spectra for different  $T<0.37\,T\sb{cr}$  (see Tab.\,\ref{t:1} and  \Fig{f:1}a), are in good agreement with the LNV prediction\,\eqref{LNV-a} with  a single fitting parameter  $ b\approx 0.5$ that replaces the factor $\frac54$ in \Eq{LNV-b}.
 \item At  $T\approx 0.37\,T\sb{cr}$  (corresponding to $\O=0.9$ in our case) we observed a critical energy spectrum $E\sb{cr}\propto 1/k^3$.
 \item The numerically observed supercritical energy spectra at $T>0.37\,T\sb{cr}$ exhibit a scale-invariant behavior $E(k)\propto k^{-x}$, Eq.\,\eqref{28A} with the scaling exponent $x>3$ that gradually increases with the temperature and reaches the value  $x\sim 9 $ for $T\approx 0.72\,T\sb{cr}$.

  \item Relaxing the assumption of locality by using integral closure for the energy flux\,\eqref{24},  we confirmed analytically  the scale-invariant spectrum    $ E(k)\propto k^{-x}$, Eq.\,\eqref{28A} with the variable scaling exponent $x$ that depends on the temperature  in a qualitative agreement with the DNS observation.

\item In the near-critical regimes we observed significant increase in turbulent fluctuations of superfluid velocity and vorticity at small scales, typical for intermittency.
 \end{description}

\section{\label{s:He3}Analytic Discussion of the Statistics  of $\bm ^3$He turbulence}

\subsection{\label{ss:HVBK} Gradually damped HVBK-equations for   superfluid $^3$He-B turbulence}
Large scale turbulence in superfluid $^3$He can be described by the Landau-Tisza two-fluid model in which the interpenetrating normal and superfluid components have densities $\rho\sb n$,  $\rho\sb s$ and  velocity fields $\B u\sb n(\B r,t)$, $\B u\sb s(\B r,t)$, respectively. The gradually damped version of the coarse-grained  HVBK  equations \cite{PRB} for incompressible motions of superfluids with constant densities has the form of two Navier-Stokes equations supplemented by mutual friction:
\begin{subequations}\label{NSE} \begin{eqnarray}   \label{NSEs} 
 && \hskip -1.3cm \frac{\p \,\B u\sb s}{\p t}+ (\B u\sb s\* \B
\nabla) \B u\sb s   - \frac 1{\rho\sb s }\B \nabla p\sb s  =\nu\sb s\,  \Delta \B u\sb s   + \B f \sb {ns}\,, 
 \\  \label{NSEn}
&& \hskip -1.3cm   \frac{\p \,\B u\sb n}{\p t}+  ( \B u\sb n \* \B \nabla)\B u\sb n  - \frac 1{\rho\sb n }\B \nabla p\sb n    = \nu\sb n\,  \Delta \B
u\sb n  \B -  \frac{\rho\sb s}{\rho\sb n}\B f \sb {ns}\, ,\\ \nn
 && \hskip -1.3cm  p\sb n =\frac{\rho\sb n}{\rho }[p+\frac{\rho\sb s}2|\B u\sb s-\B u\sb  n|^2]\, ,
   p\sb s =  \frac{\rho\sb s}{\rho }[p-\frac{\rho\sb n}2|\B u\sb s-\B u\sb n|^2]\, ,\\ \label{1e}
   \B f\sb {ns}&\simeq&  \a(T)\, \Omega\Sb T\,(\B  u \sb n-\B  u \sb s ) \, .
\end{eqnarray}\end{subequations} 
Here   $p\sb n$,  $p\sb s$   are  the pressures  of the normal and the superfluid components.
$\rho\equiv  \rho\sb s+\rho\sb n$  is the total density, $\nu\sb n$ is the kinematic viscosity of normal fluid component.  The dissipative term with the Vinen's effective superfluid viscosity  $\nu\sb s$  was added in \Ref{He4} to account for the energy dissipation at the intervortex scale $\ell$ due to vortex reconnections and similar effects.  A qualitative  estimate of the effective viscosity $\nu\sb s \simeq \alpha \kappa \rho\sb s/ \rho$ follows from a model of a random vortex tangle moving in a quiescent normal component\,\cite{He4}.

The approximate \Eq{1e} for the mutual friction force
$\B f\sb {ns}$ was suggested in \Ref{LNV}. It involves  the temperature dependent dimensionless mutual friction parameters $\alpha(T)$ and  rms superfluid turbulent vorticity $\Omega\Sb T$. In isotropic turbulence
\begin{equation}\label{ot}
 \Omega\Sb T ^2 \=  \< |\B \o|^2\>\approx 2 \int  k^2  E\sb s(k)d k\,,
\end{equation}
 where $E\sb s(k)$ is the one-dimensional (1D) energy
spectrum, normalized such that the total energy density per
unit mass $\C E\sb s =\int  E\sb s (k)\, d k$.

Note that in \Eq{NSE} we did not account for the reactive part of the mutual friction\,\cite{Sonin}, proportional to another temperature dependent parameter $\a'$. As was shown in \Ref{Finne}, this force leads to a renormalization of the nonlinear terms  in \Eq{NSEs} by a factor  $(1-\a')$. Dividing \Eq{NSEs} by this factor, we see that (besides the renormalization of time) we get also the renormalization
of $\alpha\Rightarrow \~ \alpha= \a/(1-\a')$ in \Eq{1e}, which now reads:

\begin{equation}\label{tildeOm}
\B f\sb {ns} \simeq  \Omega \,(\B  u \sb n-\B  u \sb s ) \,,
   \   \Omega = \~ \a(T)\, \Omega\Sb T \,, \  \~ \alpha= \a/(1-\a')\ .
   \ee

Ideally, the turbulent vorticity $\Omega\Sb T$  should be calculated self-consistently, at each time step. However we use a simplified version, by first solving \Eqs{NSE} with some value of $\O$, then calculating $\Omega\Sb T$ by \Eq{ot} with the observed $ E\sb s(k)$ and finally  finding $\alpha\Sp{DNS}=\Omega/\Omega\Sb T$.  After that we identify the temperature to which  the particular simulation corresponds by comparing with known experimental values $\a(T)=\a\Sp{DNS}$.  We have verified that in the present range of parameters, simulations with a constant value of  $\Omega$ and  self-consistent simulations give similar results.

\subsection{\label{ss:defs}  Statistical description of   space-homogeneous, isotropic turbulence of superfluid $^3$He}
\subsubsection{\label{ss:defs1}Definition of 1-D energy spectra and cross-correlations}
 Traditionally one describes the energy distribution over scales in a space-homogeneous, isotropic case using the one-dimensional (1D) energy spectrum $E(k)$, defined by \Eq{def1a}. To clarify this definition we need to recall some well known  relationships.

Fourier transforms are defined with the following normalization:
\begin{subequations}\label{FT}
 \begin{eqnarray}\label{FTa}
 \B u\sb{n,s}(\B r,t)& \= &   \int   \frac{ d \B k}{(2\pi)^3}   \,   \~{\B
u}\sb{n,s}(\B k,t) \exp(i \B k\* \B r )\,, \\ \label{FTc}
  \~{\B u}\sb{n,s}(\B k,t)  &=&    \int    d \B r ~   \B
u\sb{n,s}(\B r,t) \exp(-i\B k\* \B r)\ .~~~~~~~
\end{eqnarray} 
\end{subequations}

Next we define the simultaneous correlations and cross-correlations in $\B k$-representation, [proportional to $\d(\B k + \B q )$ and $\d(\B k + \B q +\B p)$ due to the space homogeneity]:
\begin{subequations}\label{corr}
\begin{eqnarray}\label{corr-nn}
&&\< \~{\B u}\sb n(\B k,t)\*\~{\B u}  \sb n(\B q ,t)  \>= (2\pi)^3 F\sb{nn}(\B k)\, \d(\B k+\B q )\,,~~~ \\
\label{corr-ss}
&& \< \~{\B u}\sb s(\B k,t)\*\~{\B u}  \sb s(\B q ,t)  \>=(2\pi)^3 F\sb{ss}(\B k)\, \d(\B k+\B q )\,, \\ \label{Corr-ns}
&&\< \~{\B u}\sb n(\B k,t)\*\~{\B u}  \sb s(\B q ,t)  \>= (2\pi)^3 F\sb{ns}(\B k)\, \d(\B k+\B q )\,,
 \\ \nn
&&\< \~{  u}^{\,\xi} \sb s (\B k,t)\, \~{ u}^{\, \beta} \sb s(\B q,t)\, \~{u}^{\, \gamma}  \sb s(\B p ,t)  \> \\ \label{Corr-sss}
&=& (2\pi)^3 F ^{\xi \beta \gamma }\sb{sss}(\B k,\B q, \B p)\, \d(\B k+\B q +\B p)\ .
\end{eqnarray}\end{subequations}

In the isotropic  case   the correlations $F\sb{nn}$,  $F\sb{ss}$ and $F\sb{ns}$ become  independent of the direction of $\B k$, being functions of the wavenumber $k$ only.     This allows us to introduce the one-dimensional energy spectra $  E\sb s$, $  E \sb n$ and the cross-correlation $  E\sb{ns}$ as follows:
 \begin{eqnarray}\nn 
 E\sb{n}(k)&=& \frac{k^2}{2\pi^2}F\sb{nn}(k)\,, \quad   E\sb{s}(k)= \frac{k^2}{2\pi^2}F\sb{ss}(k)\,,\\ \label{def1a}
  E\sb{ns}(k)&\=& \frac{k^2}{2\pi^2}F\sb{ns}(k)\ .
\end{eqnarray}

\subsubsection{\label{ss:defs2}Energy balance equation}
To derive the energy balance equation for $E\sb s(k,t)$ we first need to Fourier transform \Eq{NSEs} to get the equation for $\~{\B u}\sb s(\B k,t)$. Next, using \Eq{corr-ss} and \Eq{def1a},
     we arrive to the required balance equation:
  \begin{subequations}\label{BAL} \begin{eqnarray}\label{BALa}
  && \frac{\partial E\sb s (k)}{ \partial t}+  \mbox{Tr}(k)+ \mbox{D}_\nu(k) + \mbox{D}_\alpha(k)=0\,, \\
  && \mbox{D}_\nu = 2\, \nu\sb s k^2  E\sb s (k)\,, \  \mbox{D}_\alpha = 2\, \Omega \big[ E\sb{s}(k)-E\sb {ns} (k)  \big]\ . ~~~~
   \end{eqnarray}
 Here D$_\nu$ describes the energy dissipation, caused by the effective viscosity. The term  D$_\alpha$ is responsible for the energy dissipation by the mutual friction with the characteristic frequency $\Omega$ given by \Eqs{1e} and  \eqref{ot}.

The energy transfer term Tr$(k)$ in \Eq{BALa} originates from the nonlinear terms in the HVBK \Eqs{NSEs} and has the same form as in classical turbulence (see, e.g. \Refs{LP-95,LP-2}):
  \begin{eqnarray}\nn
 \mbox{Tr}(\B k)&=& 2\, \mbox{Re}\Big\{\int V^{\xi\beta\gamma}(\B k,\B q,\B p)\, F^{\xi\beta\gamma}(\B k,\B q,\B p)\\ \label{genA}
 && \times \delta(\B k+\B q+\B p)\frac{d^3 q \, d^3 p}{(2\pi)^6} \, \Big\}\,, \\ \nn
 V^{\xi\beta\gamma}(\B k,\B q,\B p)&=& i  \Big ( \delta _{\xi \xi'}- \frac{ k^\xi k^{\xi'}}{k^2}   \Big )\\ \label{genB}
  && \times \Big( k^\beta \delta_{\xi ' \gamma} + k^\gamma \delta _{\xi' \beta} \Big )\ .
 \end{eqnarray}
  Importantly, Tr$(k)$ preserves the total turbulent kinetic energy: $\displaystyle \int_0 ^k\mbox{Tr}(k')dk'=0$ and therefore can be written in the divergent form:
\begin{equation}\label{BALb}
\mbox{Tr}(k)= \frac{\partial\,  \ve (k)}{d k}\,,
\end{equation}\end{subequations}
where $\ve(k)$ is the energy flux over scales.

 \subsection{\label{ss:integral} Supercritical energy spectra}

\subsubsection{\label{sss:int} LNR integral closure}
To relax the assumption of the local energy transfer in deriving the supercritical superfluid energy spectrum, we use the integral closure, introduced by L'vov, Nazarenko and Rudenko\cite{LNR}(LNR).
  The main approximation in this closure is the presentation of the third order velocity correlation function $F^{\xi\beta\gamma}\sb{sss}$ in \Eq{genA} as a product of the vertex $V$, \Eq{genB},  two second order correlations $F\sb{ss}(k_j)$, \Eq{corr-ss}, and  response (Green's) functions.
  This closure is  widely used in analytic theories of classical turbulence, for example in the Eddy-damped quasinormal Markovian closure  (EDQNM) (see, e.g. books \Ref{Fri,DKS} ).   Keeping in mind the uncontrolled character of this approximation, LNR further simplified the resulting approximation for isotropic turbulence by replacing  $d^3 q
\, d^3 p \, \d^3( \B k +\B  q+ \B  p) $ in \Eq{genA} with 3-dimensional vectors
$\B  k$,   $\B  q$, and $\B p$~~ by~ $ q^2 d q \, p^2 d   p
 \  \d (  k +  q+   p)/ (k^2+q^2+p^2)$ with
  one-dimensional vectors  $  k$,   $  q$, and $  p$  varying
in the interval $(-\infty, + \infty)$. The next simplification is the replacement of the interaction amplitude $V^{\xi\b\g}(  \B k, \B q, \B p)$, \Eq{genB}  by its  scalar  version
  $ (i  k)$.  The resulting LNR closure can be written as follows:
\begin{eqnarray}\label{24}
   &&\mbox{Tr}(k)=\frac { A_1\, k^3}{2\pi^2}
   \int  _{ -\infty}^\infty
  \frac{q^2 d   q \, p^2 d   p
 \, \d (  k +  q+   p)}{2\pi \, (k^2+q^2+p^2)}\br
&&\hskip -.5cm   \times \frac{       k\, F\sb {ss}(|q|)  F\sb {ss}(|p|) +
    q\, F\sb {ss}(|k|)  F\sb {ss}(|p|) +    p\, F\sb{ss}(|q|) F\sb {ss}(|k|)    }{\Gamma(|k|)+\Gamma(|q|)+\Gamma(|p|)}  .
\eea 
Here $A_1$ is a dimensionless parameter of the order of unity and $\displaystyle \Gamma(k)$ is the typical
relaxation frequencies on the scale $k$.

The LNR model~\eq{24}  satisfies all the general closure requirements: it
conserves energy, $\int  \mbox{Tr}(k)\,dk =0$  for any  $F_k$;  Tr$(k)=0 $ for the thermodynamic equilibrium spectrum
$F_k=$const and  for the cascade K41 spectrum $ F(k)\propto |k|^{-11/3}$.
Importantly, the integrand in \Eq{24}  has the correct  asymptotic
behavior at the limits of small and large $q/k$, as required by the
sweeping-free Belinicher-L'vov representation, see \Ref{BL}. This
means that the model\,\eqref{24} adequately  reflects contributions of the
extended interaction triads and thus  can be used for the analysis of the supercritical spectra.

\subsubsection{\label{sss:super} Supercritical spectra with non-local energy transfer}
 As was shown in \Ref{Kh15},  the eddy life time in  $^3$He turbulence is restricted by the mutual friction, which dominates  the dissipation due to the effective viscosity $\nu_s k^2$ and the turbulent viscosity, caused by the eddy interactions. Therefore we can safely approximate   $\Gamma(k)$ in \Eq{24} by  $\Omega$.  Omitting further the (uncontrolled) prefactors of the order of unity and
 using \Eq{def1a},  we rewrite Tr$(k)$ in \Eq{BALa} as follows
\begin{subequations}\label{balE}

\begin{eqnarray}\label{25}
\mbox{Tr}(k)&\simeq& - \frac {A\,k} \Omega
   \int\limits  _{ -\infty}^\infty
  \frac{  d   q \,   d   p
 \, \d (  k +  q+   p)}{k^2+q^2+p^2}\\ \nn
 &&\hskip -1.5cm \times {   \big [   k^3 \,  E\sb s(|q|)   E\sb s(|p|) +
    q^3   E\sb s(|k|)  E\sb s(|p|) +  p^3\,  E\sb s(|q|)   E\sb s(|k|)    \big ]} \ .
\end{eqnarray}

Here $A$ is uncontrolled dimensionless parameter, presumably of the order of unity.
Recall, that in $^3$He turbulence $E\sb{n}\ll E\sb s$ and $E\sb{ns}\ll E\sb s$. This allows us to simplify the mutual friction dissipation term D$_\alpha$ to the form D$_\alpha(k)\approx  2 \, \Omega E\sb s(k)$.  Hereafter we consider only superfluid component and omit the superscript ''s`` in notations.
We show below that in the supercritical regime the viscous dissipation term D$_\nu(k)$ is vanishingly small with respect to the mutual friction term D$_\alpha(k)$ and therefore can be neglected in the balance \Eq{BALa}. Thus, in the stationary case \Eq{BALa} can be presented in a simple form:
 \begin{equation}
 \mbox{Tr}(k) +2\,\Omega \,  E(k)=0\ .
 \end{equation} \end{subequations}
The integral\,\eqref{25} diverges in the regions $q\ll k$ or $p\ll k$. For these wavenumbers it can be approximated as:

\begin{eqnarray}    \label{26}
 \mbox{Tr}(k)   &\simeq & - \frac {A\, k}  \Omega
   \int\limits  _{ -\infty}^\infty
    E(|q|) \Psi(k,q)  d   q\,,
   \\ \nn
 \Psi(k,q)&=& \frac{       k^3 \,  E(|k+q|)
     - (k+q)^3\,  E(|k|)     }{k^2+q^2+(k+q)^2}\ .
\end{eqnarray}

One sees that for $q=0$ $\Psi(k,0)=0$ and the term which is linear in $q$ in the expansion does not contribute to the integral\,\eqref{26}. Therefore the main contribution to this integral in the region $q\ll k$ originates from the second term of the expansion:
\begin{eqnarray}\label{27}
\Psi(k,q)\simeq \frac {q^2}{2}\frac{\partial ^2 \Psi}{\partial  q^2}\Big |_{q=0}=\frac {q^2}{2}\Big[\frac k2  E^{\prime\prime}(k) -  E^{\prime}(k)\Big]\, .
\end{eqnarray}
Here $^{\prime}$ indicates the   derivative with respect to $k$.
Now the energy balance \Eqs{balE} can be simplified as follows:

\begin{equation}\label{res1A}
  A \, \O\Sb T^2 k \Big [ \frac k2  E^{\prime\prime}(k) -  E^{\prime}(k)\Big ] =4\,\O^2\,   E(k)\,,
  \ee
where $\Omega\Sb T$ is given by \Eq{ot}.
Equation\,\eqref{res1A} has the scale invariant solutions
\begin{subequations}\label{28}
\begin{equation}\label{28A}
 E(k)\propto k^{-x}\,,
\ee  with
\begin{equation}\label{28B}
  A \, \Omega\Sb {T}^2\,  x(x-1) = 8\, \O^2\ .
\ee
The whole approach is valid if the main contribution to the integral\,\eqref{ot} comes from the region $q\ll k\sb{max}$, i.e for supercritical cases with $x>3$. With logarithmic accuracy we can also include the critical case with $x=3$. This allows us to estimate the new critical value of $\O$ for supercritical regimes (with $x>3$):
\begin{equation}\label{28D}
\~ \Omega\sb{cr}=\Omega \Sb T\sqrt{3 A}/2\ .
\end{equation}
Now we can rewrite \Eq{28B} as:
\begin{equation}\label{28C}
  x(x-1) = 6 \, (\Omega^\ddag)^2\,, \quad \Omega^\ddag \= \Omega/\~\Omega \sb{cr} \,, \quad x>3\ .
\ee
\end{subequations}
We thus conclude that for the integral closure\,\eqref{25} that takes into account the long-distance energy transfer in $k$-space, the supercritical spectra do not terminate at some final value of $k$ [as with the algebraic  closure\,\eqref{alg}], but behave like $ E(k)\propto k^{-x}$ with a scaling exponent $x>3$ that increases with the supercriticality $\Omega^\ddag$.

\subsection{\label{sss:struct} Relations between structure functions and energy spectra}

\paragraph{Velocity structure function $S_2(r)$ vs   $ E(k)$.}

Consider   full 2$\sp{nd}$-order velocity structure function
  \begin{subequations}\label{8}
 \begin{equation}\label{8a}
  S_2(r)\=  \< |\B v (\B r + \B R)- \B v (\B R)|^2\>\,,
   \end{equation}

which is related to the 3D energy spectrum $ F(k)$ as follows:

   \begin{eqnarray}\label{8b1}
   S_2(r)&=&\int \frac{d^3 k}{(2\pi)^3} \, |1-\exp(i \B k \cdot \B r)|^2 F (k)\\ \nn
  &=& 2 \int \frac{d^3 k}{(2\pi)^3}   \big [ 1-\cos( \B k \cdot \B r)\big ]  F (k)\ .
    \end{eqnarray}

    \end{subequations}

  In spherical coordinates:
  \begin{eqnarray}\label{9}
  S_2(r)&=& 2 \int   E(k)\Big [ 1 - \frac{\sin(kr)}{kr}\Big ] d k\ .
  \end{eqnarray}

 Let us analyze convergence of this integral for scale-invariant spectra $ E(k)\propto k^{-x}$.  In the ultraviolet (UV) region (for $k\,r\gg 1$) the oscillating term ($\propto \sin(k\,r)$) can be neglected and the integral\,\eqref{9} converges if $x>1$.    In the infrared (IR) region (for  small $k\, \ll 1$)
 \begin{equation}\label{lim}
 [1-\sin(k\,r)/(k\, r)]\simeq (k\,r)^2/6
  \ee
  and the integral\,\eqref{9} converges if $x<3$. We conclude that for the integral\,\eqref{9} the window of convergence (more often is referred to as the \emph{locality window}) is:
 \begin{subequations}\label{S2}
 \begin{equation}\label{S2a}
 1< x < 3\,, \quad \mbox{Locality window for $S_2$ integral.}
 \ee
In this window, the leading contribution to the integral\,\eqref{9} comes from the region $k\,r\sim 1$ and
\begin{equation}\label{S2b}
S_2(r)\propto r^y\,, \quad y= x-1\ .
\ee\end{subequations}
This is a well know relationship. For example, for the K41 spectrum with $x=5/3$  (which is inside the locality window\,\eqref{S2a}) $y=2/3$ .

We conclude that subcritical spectra, (which  in the finite-$k$ interval can be  approximated as  $ E(k)\propto k^{-x}$ with $ \frac53 \leq x \leq 3$)  are local and we can use for the estimate of the $S_2$ the scaling relation\,\eqref{S2b}. We also see that when exponent $x$ approaches the critical value $x=3$, the $S_2$ scaling approaches the viscous limit with $y=2$. For $x=3$,   $S_2(r)\propto r^2$ with logarithmic corrections, not detectable with our resolution.

 In the supercritical region ($x>3$), the $S_2$-integral\eqref{9} formally IR-diverges and the integration region has to be restricted from below by  some $k_0$, similarly to the integral\,\eqref{ot}. Together with \Eq{lim}, this gives the viscous behavior for any $x>3$:
 \begin{equation}\label{17}
 S_2(r)\simeq (r\,\Omega\Sb  T)^2/6\ .
 \ee

\begin{table*}[t]
  \caption{\label{t:1} Parameters used in the simulations by columns:
  (\# 1) $\O$ determines the mutual friction by  \Eqs{1e} and \eqref{ot};
  (\# 2) $\nu_s$: the effective viscosity of the superfluid component;
  (\# 3) $u\sp s_{\rm rms}$: the rms velocity of the superfluid component;
  (\# 4) $Re\sp s_\lambda= u\sp s_{\rm rms} \lambda/\nu_s$: the Taylor-microscale Reynolds number,
 where $\displaystyle \lambda = \frac{2\pi}{L}\sqrt{\frac{\langle u^2\rangle}{\langle \omega^2 \rangle}}$ is the Taylor  microscale;
 (\# 5) $\varepsilon\sp s_\nu$: the mean energy dissipation rate for the superfluid component due to viscosity;
 (\# 6) $\varepsilon\sp s_{\rm tot}$: total mean energy dissipation rate for the superfluid component;
 (\#7) $\eta_s=\sqrt{2}\nu_s/u\sp s_{\rm rms}$;
 (\# 8) $T\sp s_0=L/u\sp s\sb {rms}$: large-eddy-turnover time.
 The temperature dependence of $\~ a$ is taken from Ref.\cite{Bevan} (see \Fig{f:8}). In all simulations: the number of collocation points along each axis is $N=1024$;
 the size of the periodic box is $L=2\pi$;
 the kinematic viscosity of the normal component is $\nu_n=10$;
 the range of forced wavenumbers $k^{\varphi}=[0.5 , 1.5]$.
 The values for the critical value of  $\Omega\sb{cr} \approx \Omega=0.9$ [row (\#5)] are emphasized. Runs (\#1-\#4) correspond to the subcritical regime, (\#6-\#8) to the supercritical regime.  }

\begin{tabular*}{\linewidth}{@{\extracolsep{\fill} } c     c   c c c c c  c  c   c  c  c    c  c  c c  }
    \hline
    \hline
    \#  &1 &2&3&4&5&6&7&8&9&10&11&12&13 &14 & 15 \\
   \#  &  $\Omega$     & $\nu_s$    & $u^s_{\rm rms}$ & $Re^s_\lambda $
   &$\varepsilon^s_\nu$ &$\varepsilon^s_{\rm tot}$  & $\eta_s $  & $T_0^s$  & $\Omega\sb{cr}$&$\~ \Omega\sb{cr}\approx $ & $ \Omega^\dag= $&  $\Omega ^\ddag=$ &  $\Omega\Sb T$ &$1/\~\alpha(T)$ & $T/T\sb c$ \\
   & \Eq{tildeOm}&  $\times 10^4$  &    &    &     &    & $\times 10^4$ &   & \Eq{Om-cr-1} & $  0.18\,  \Omega\Sb T$  &$\Omega/\Omega\sb {cr}$ &  $\Omega/\~ \Omega\sb {cr}$  &  \Eq{ot} & \Eq{tildeOm}&   \\
    \hline
   1 &  $0$ & $5$   &$1.14$ & $0$   &$4.6$ &  $590$  &   $4.95$ &  $4.95$ & $1.14$ & 17.7 &$0$ & $-$ & 100  & $\infty$ & 0 \\
   2  &  $0.25$& $5$&$0.89$ & $0.28$&$3.3$ &  $750$  &  $0.85$ & $3.57$ & $0.89$ & 7.4& $0.28$ & $-$  & 41 &164  & 0.19 \\
   3  &  $0.5$ & $1$& $0.95$&$0.53$ &$3.2$ &  $2600$ &    $0.34$ & $5.5$ & $0.95$ & 10.4& $0.53$  & $-$ & 21 &42  & 0.27 \\
   4 &  $0.7$ & $1$ & $0.81$&$0.86$ &$2.6$ &  $7500$  &    $0.015$ & $4.38$ & $0.81$ & 2.2 &  $0.86$ & $-$ & 10.4 &15  & 0.32  \\ \hline
   5 &  $0.9$ & $1$ & $0.79$&$1.13$ &$2.5$ &  $16000$ &  $0.0028$ & $5.1$& $0.79$ & 0.9& $-$ & 1.0 & 6.3 & 7 & 0.37 \\ \hline
   6  &  $1.1$ & $1$& $0.75$&$1.46$ &$2.3$ &  $23000$ &    $0.001$ & $5.2$  & $0.75$ &0.55 & $-$ & 2.0 &3.3  &3 & 0.39 \\
   7  &  $2.5$ & $1$& $0.57$&$4.42$ &$1.6$ &  $18000$ & $0.0004$ &  $5.53$ & $0.57$ &0.3& $-$ & 8.4& 2.0 & 0.8 & 0.59  \\
   8  &  $5$ & $1$  & $0.4$ & $12.1$&$1.2$ &  $14000$  &    $0.0002$ &  $5.2$ & $0.4$ & 0.21& $- $ & 24& 1.4 &0.3  & 0.72  \\
    \hline
    \hline
  \end{tabular*}
\end{table*}

\paragraph{Vorticity  structure function $T_2(r)$ vs $ E(k)$.}

Consider now  2$\sp{nd}$-order vorticity  structure function
 \begin{subequations}\label{vort}
 \begin{equation}\label{vortA}
    T_2(r)\=  \< |\B \o (\B r + \B R)- \B \o (\B R)|^2\>\,,
   \end{equation}
 which is related to the 3D energy spectrum  $ F(k)$ as follows:
   \begin{eqnarray}\label{8b2}
 T_2(r)&=&\int \frac{d^3 k}{(2\pi)^3} \, |1-\exp(i \B k \cdot \B r)|^2 k^2  F (k)\\ \nn
   &=& 2   \int k^2   E(k)\Big [ 1 - \frac{\sin(kr)}{kr}\Big ] d k\ .
    \end{eqnarray}\end{subequations}

By analogy, we can immediately find the locality window of this  integral
 \begin{subequations}\label{T2}
 \begin{equation}\label{T2a}
3< x < 5\,, \quad \mbox{Locality window of $T_2$ integral.}
 \ee
Within this window
 \begin{equation}\label{T2b}
 T_2(r)\propto r^z\,, \quad z=x-3\ .
 \ee
It is also clear that for $x>5$ the scaling of $T_2(r)$ takes the form
\begin{equation}\label{T2c}
T_2(r) \simeq \frac{r^2} 3 \int q^4  E(q) dq  \sim r^2 \, \Omega\Sb  T^2\,  k_0^2\ .
\ee\end{subequations}

\section{\label{s:DNS} Statistics  of $\bm ^3$He turbulence: ~~~    DNS results and their analysis}
\subsection{\label{ss:procedure} Numerical procedure}

We carried out a series of DNSs of Eqs.~(\ref{NSEs}) and (\ref{NSEn}) using a
fully de-aliased pseudospectral code up to  $1024^3$ collocation points in a triply
periodic domain of size $L=2\pi$. In the numerical evolution, to get to a stationary state we further
 stir the velocity field of the normal and superfluid components with a random Gaussian forcing:

\begin{equation}\label{force}\langle {\bm \varphi}_u({\bm k},t)\cdot {\bm \varphi}_u^*({\bm q},t') \rangle =\Phi(k) \delta( {\bm k}-{\bm q}) \delta(t-t')
\widehat P({\bm k})\,,
 \end{equation}
 where $\widehat P({\bm k})$ is a projector assuring incompressibility
and $\Phi(k)=\Phi_0 k^{-3}$; the forcing amplitude $\Phi_0$ is nonzero only in a given band of Fourier modes:
 $ k^{\varphi} \in [0.5,1.5]$
 . Time integration is performed with a 2nd order
Adams-Bashforth
scheme with viscous term exactly integrated. The parameters of the
Eulerian dynamics for all runs are reported in
Table~\ref{t:1}.

\begin{figure*}
 \begin{tabular}{cc}
 a  &
 b  \\
 \includegraphics[scale=0.43 ]{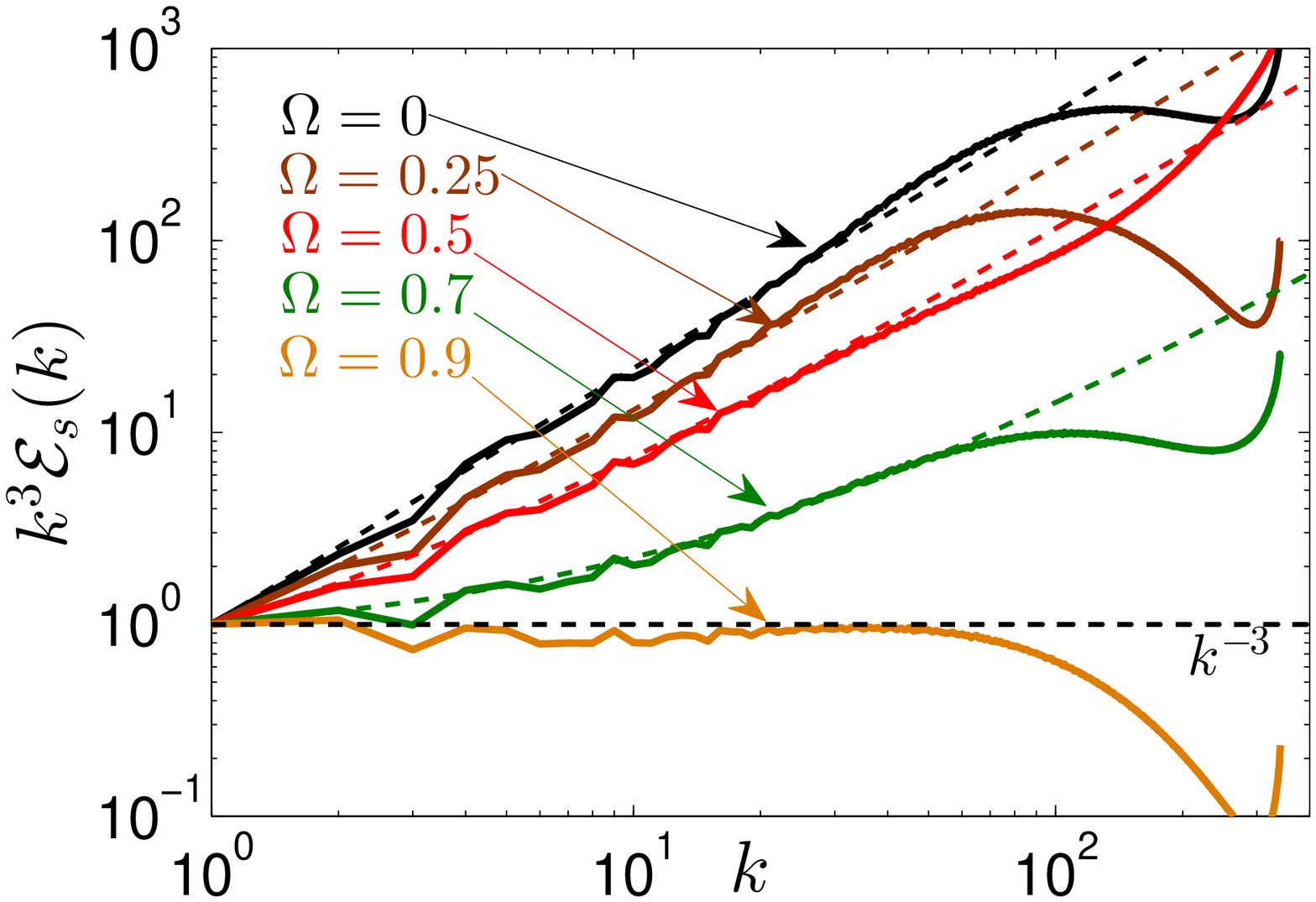}&
 \includegraphics[scale=0.43 ]{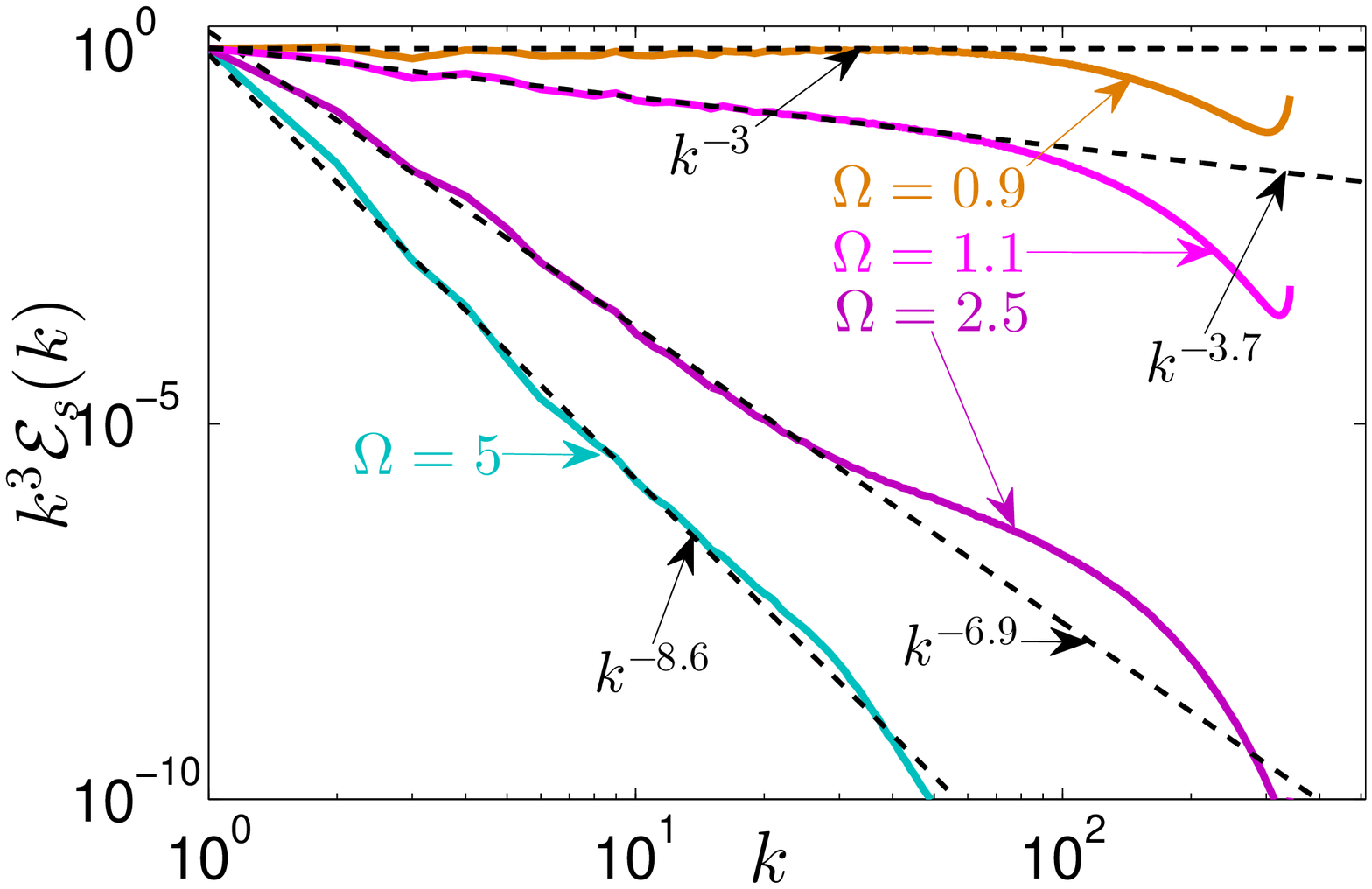}\\
  \end{tabular}
  \caption{\label{f:1} The normalized energy spectra $\C E_s(k)=E(k)/E_0$ compensated by $k^3$: subcritical   [Panel (a)] and supercritical [Panel (b)] (solid lines) for different values of $\Omega$. The critical spectrum (with $\Omega=0.9$) is shown in both panels. The dashed lines in Panel (a) are the LNV-prediction\,\eqref{LNV-a} for the subcritical spectra  with  one fitting parameter in Eq.\eqref{Om-cr-1} ($b=0.5$) for all $\Omega< 0.9$. The horizontal dashed lines in both panels  show the critical spectrum. Other dashed lines in Panel (b) represent the  scale-invariant spectra\,\eqref{28A} with an $\Omega$-dependent exponent $x$.  }
\end{figure*}


\begin{figure*}
\begin{tabular}{cc}
(a) & (b)    \\
\includegraphics[scale=0.4]{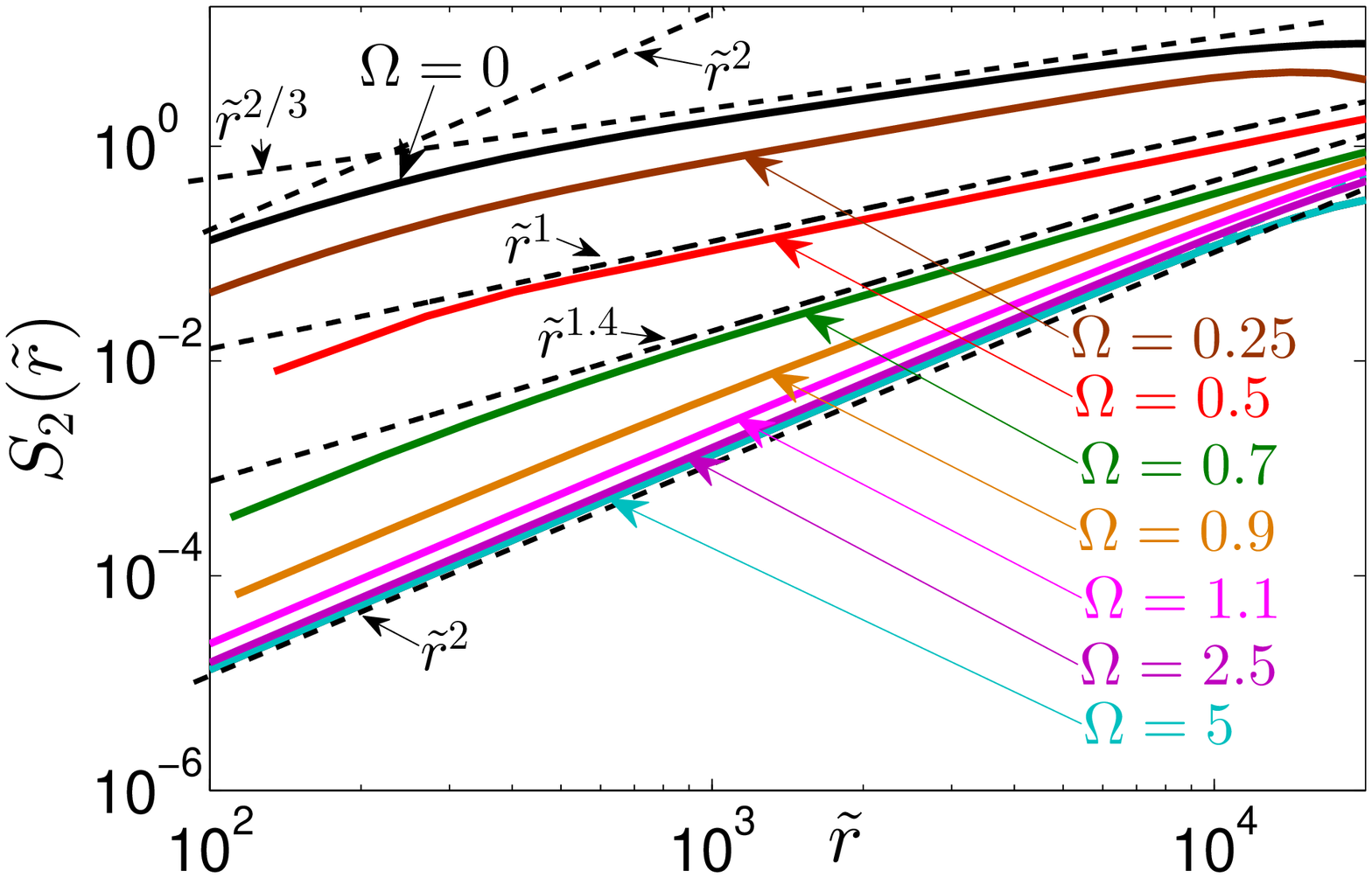}&
\includegraphics[scale=0.4]{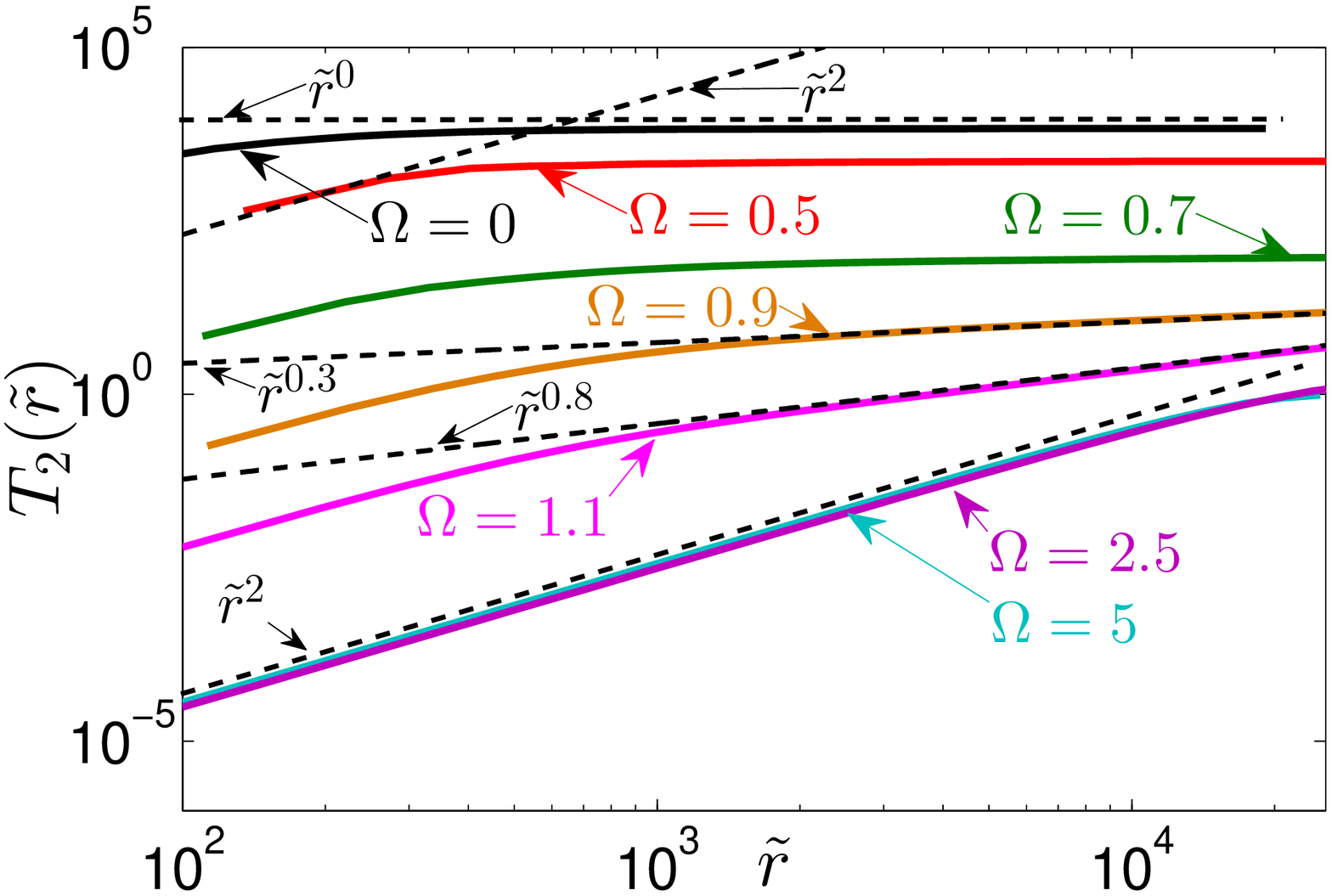}\\
(c) & (d)    \\
\includegraphics[scale=0.4]{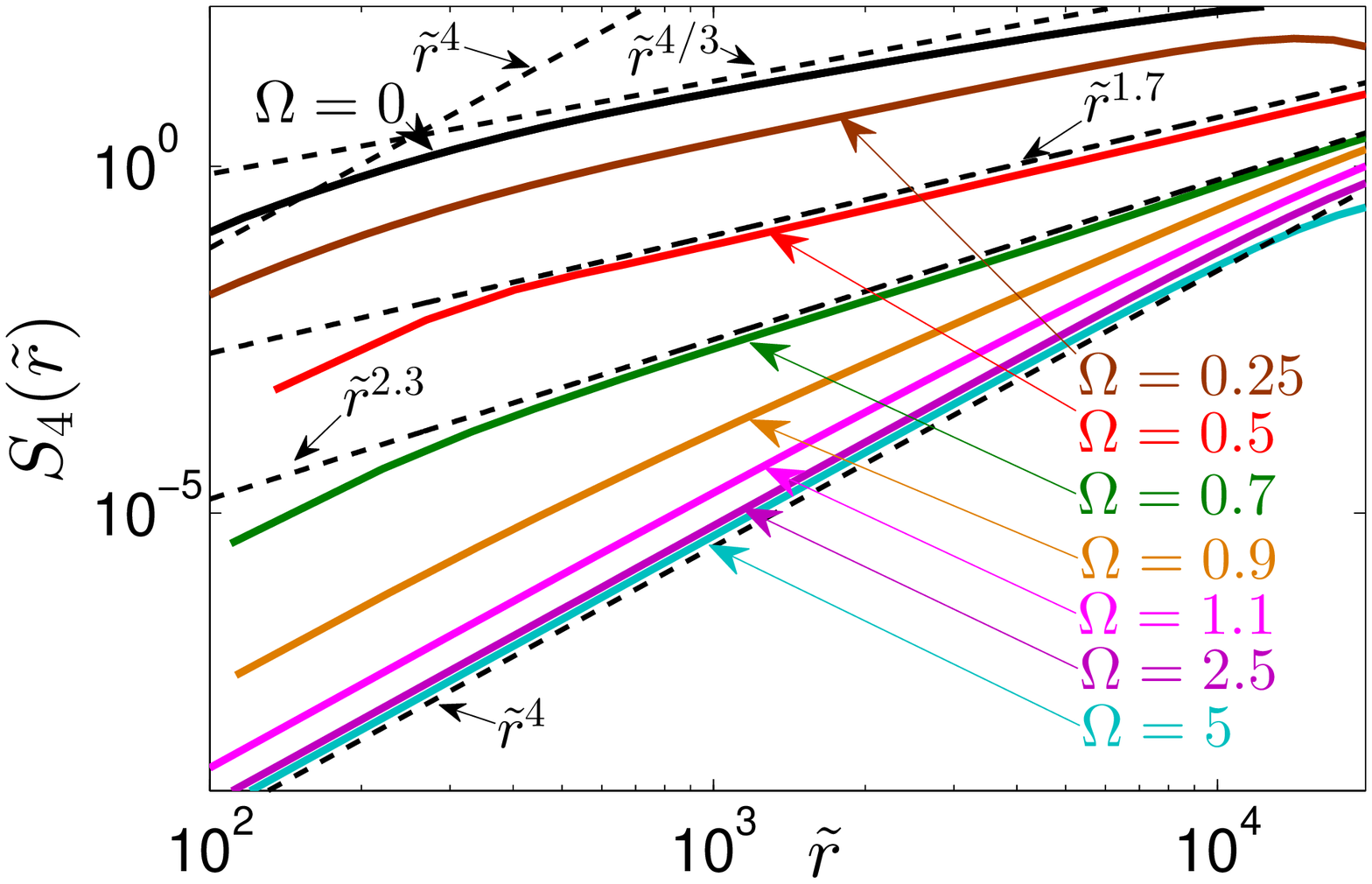}&
\includegraphics[scale=0.4]{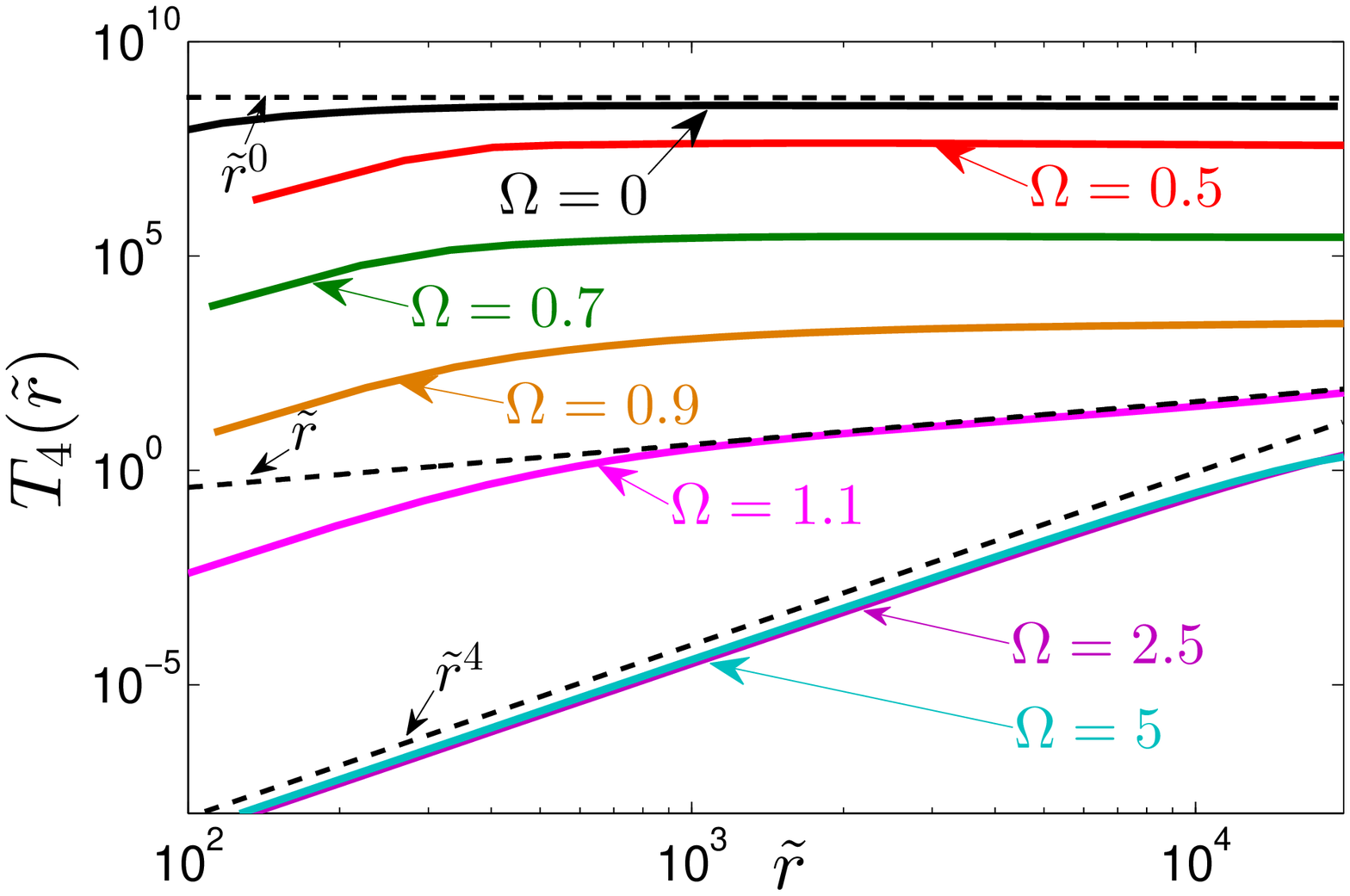}\\
\end{tabular}
  \caption{\label{f:3}Color online. The second and forth-order velocity  $S_2(\tilde r)$ and  $S_4(\tilde r)$ [Panels (a),(c)]  and vorticity $T_2(\tilde r)$ and $T_4(\tilde r)$ [Panels (b),(d)] structure functions for different $\O$. The straight dashed lines with the estimates of the apparent scaling exponents serve to guide the eye only.}

\end{figure*}

\subsection{\label{ss:energy}Energy spectra}
\subsubsection{Critical spectrum}
The numerical energy spectra are  shown in \Fig{f:1}.
As was predicted in \Ref{LNV}, at some particular ``critical" value of the mutual friction (value of $\O=\Omega\sb{cr}$ in our current notations) there exists the self-similar balance between the energy flux and the mutual-friction energy dissipation, that leads to the scale-invariant critical spectrum  $E\sb s(k)\propto k^{-3}$, \Eq{LNV-cr}.  As one sees in \Figs{f:1}, the compensated spectrum for $\Omega = 0.9$ is almost horizontal. Therefore, in our simulations $\Omega \approx 0.9$ corresponds to the critical spectrum.\\
For $\O< \Omega\sb{cr}$ we see the \emph{subcritical} spectra,  lying above the critical one.  In this case, the energy at small $k$ is dissipated by the mutual friction and approximately $E(k)\sim k^{-3}$. For larger $k$, the $k$-independent  mutual friction dissipation can be neglected compared to the energy flux (with the inverse interaction time $\gamma(k)\sim k \sqrt {k E(k)}$) and $E(k)$ can have K41 tail with the energy flux $\ve_\infty< \ve\sb{input}$, that for even larger $k$ is dissipated  by viscosity.

\subsubsection{Subcritical  LNV  spectra}

The analytical LNV-model\,\cite{LNV} of the subcritical spectra, based on the local in $k$-space algebraical closure\,\eqref{alg},   was shortly presented  in the Introduction. It results in \Eqs{Spectra} for $ E\sb{cr}(k,\O)$ formally without explicit fitting parameter. Nevertheless, having in mind   simplification\,\eqref{1e} for the mutual friction, valid up to dimensionless factor of the order of unity and the uncontrolled character of \Eq{alg} for the energy flux, we replace in \Eq{LNV-b} the numerical factor $\frac54$ by a fitting parameter $b\approx 0.5 $. Now \begin{equation}\label{Om-cr-1}
\Omega\sb{cr}= b\, \sqrt{k_0^3 E_0}\ .
\ee

 \Fig{f:1}a compares the numerical results with the analytical LNV-spectra\,\eqref{LNV-a} with $\Omega\sb{cr}$ given by  \Eq{Om-cr-1}.

 A good agreement between DNS and analytical spectra\,\eqref{LNV-a}  (with $b\approx 0.5$)  allows us to conclude that the  algebraic LNV-model with the build-in locality of the energy transfer  adequately describes the basic physical phenomena of the subcritical regime in superfluid $^3$He turbulence.

\subsubsection{Supercritical spectra}
 According to LNV model\,\cite{LNV}, for $\O> \tilde\Omega\sb{cr}$ we expect \emph{supercritical} spectra, i.e. the energy is mainly dissipated by the mutual friction and $E\sb s(k)$ falls \emph{below} the \emph{critical} spectrum $k^{-3}$. As we pointed out, the energy transfer in this regime is not local anymore and a simple algebraic closure\,\eqref{alg} fails.  Instead, we adopted an integral closure\,\eqref{24} and predicted the scale-invariant spectra $ E\sb{s}(k)\propto  k^{-x}$, \Eq{28A}, with the exponent $x$, estimated by \Eq{28B}.  As we see in \Fig{f:1}b, the supercritical energy spectra are indeed scale-invariant over more than a decade of $k$ (decaying by $13$ decades for $\Omega=5$). The scaling exponent $x$ increases with $\Omega^\ddag= \Omega/\~\Omega\sb{cr}$ as qualitatively predicted by \Eq{28B}, although much slower. For example,  $\Omega ^\ddag\approx 2.0$ for $\Omega=1.1$,  see line (\# 6) in Tab.\,\ref{t:1}. Then \Eq{28B} gives $x\sb {model}\simeq 5.4$ instead of numerically found $x\sb{num}\simeq 3.7$. This disagreement increases with $\Omega^\ddag$.     Here we should note that the particular form\,\eqref{24} of the integral closure was chosen just for simplicity. We can use much more sophisticated  kind of a two-point integral closure, like EDQNM,\cite{Ors70} or  Kraichnan's Lagrangian-history direct interaction approximation\,\cite{Kra65}, etc. However the result will be qualitatively similar: a scale-invariant solution with the exponent $x$ that increases with $\Omega^\ddag$.

 We again conclude that the suggested model (now with the integral closure) describes qualitatively  the physics of the supercritical regime of the superfluid $^3$He turbulence with the balance between the energy flux from $k\sim k_0$ directly to a given $k\gg k_0$, [the left hand side of \Eq{res1A}], where it is dissipated by the mutual friction [the right hand side of \Eq{res1A}]. This balance equation results in the power-like law $ E\sb{sp}\propto k^{-x}$, in agreement  with the DNS results. The actual value of the exponent $x$ depends on the details of the uncontrolled integral closure. A detailed analysis of the closure problem, including contribution of next order terms in perturbation approach, and comprehensive numerical simulations would  be required  to achieve better understanding of the statistics of the supercritical regimes of superfluid $^3$He turbulence

\begin{figure*}
\begin{tabular}{cc}
(a)& (b)  \\
\includegraphics[scale=0.4]{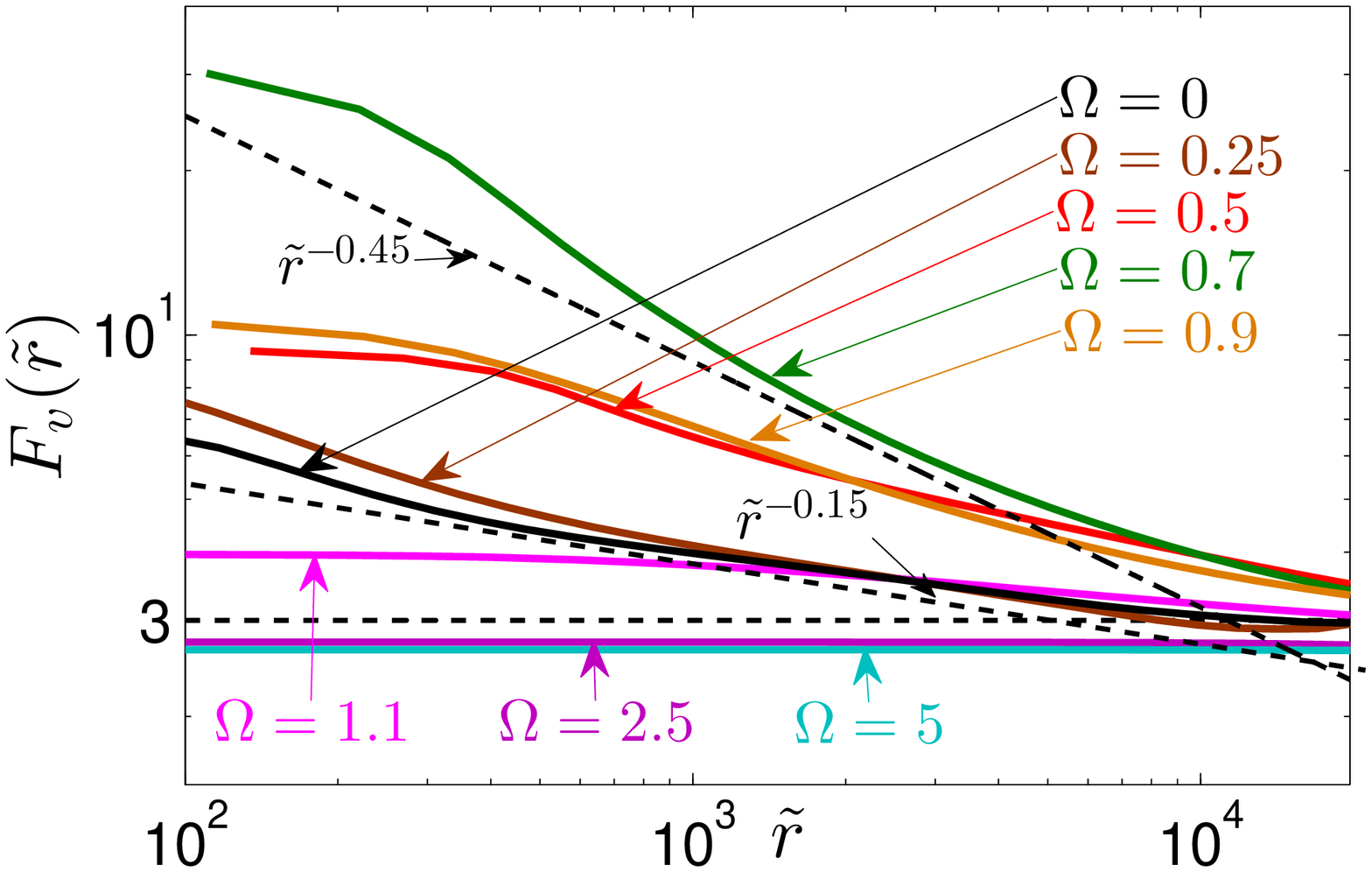}&
\includegraphics[scale=0.4]{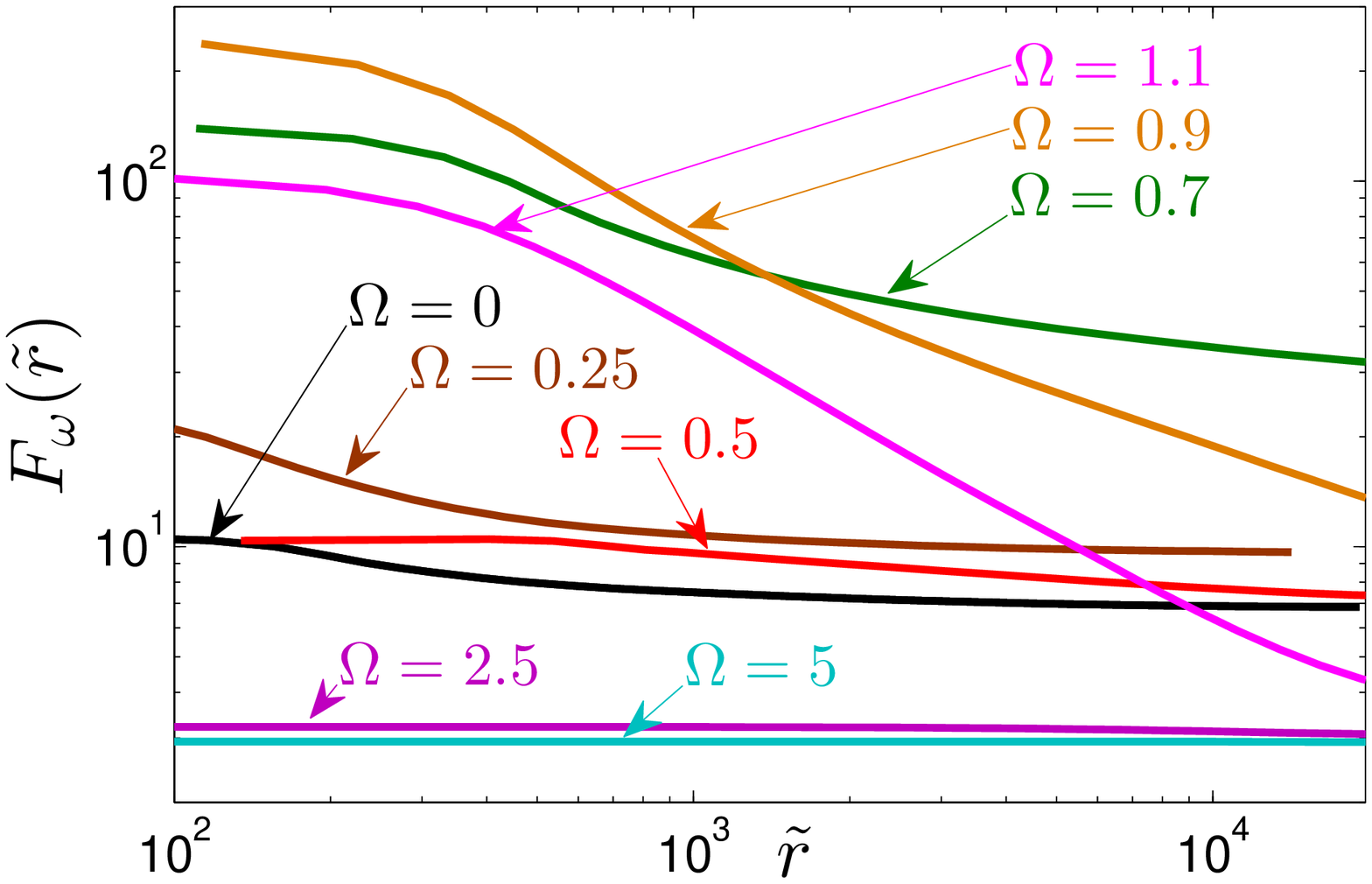}\\

(c)  & (d)  \\
\includegraphics[scale=0.4]{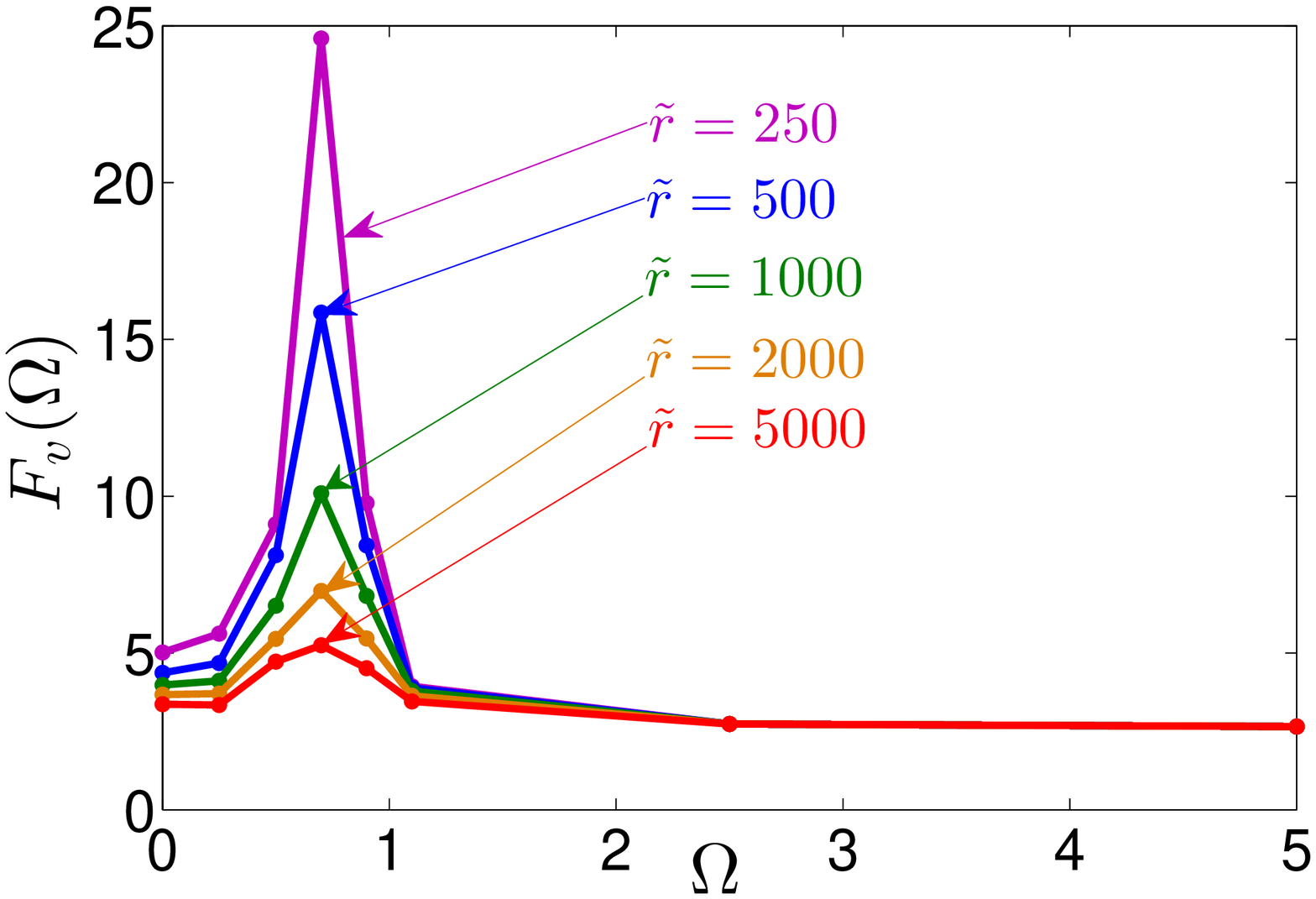}&
\includegraphics[scale=0.4]{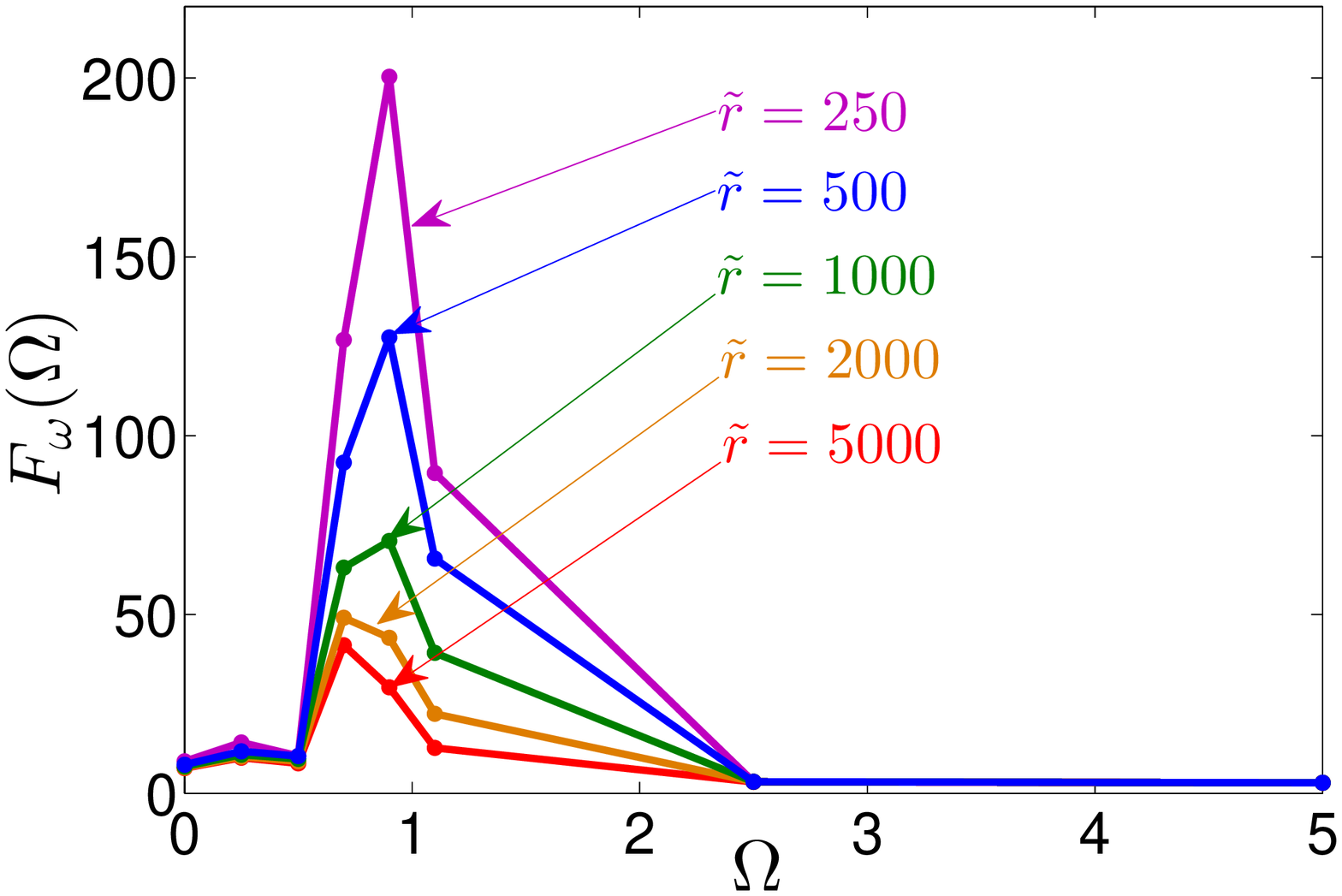}
\\
\end{tabular}
  \caption{\label{f:4}  Color online. The velocity $F\sb v( \tilde r)= S_4(\tilde r)/ S_2^2(\tilde r)$ and vorticity  $ F_\o(\tilde r)=T_4(\tilde r)/ T_2^2(\tilde r)$  flatness vs $\tilde r$ for different $\Omega$ [Panels (a) and (b)] and vs $\Omega$  for different $\tilde r$ [Panels (c) and (d)]. The straight dashed lines with the estimates of the apparent scaling exponents serve to guide the eye only. }
\end{figure*}

  \subsection{\label{ss:str}  Enhancement of intermittency in critical and subcritical regimes of superfluid $^3$He turbulence }
Current \Sec{ss:str} is devoted to the discussion of the numerically found velocity and vorticity structure functions $S_2(r)$, $S_4(r)$ and  $T_2(r)$, $T_4(r)$ and to comparison their  scaling with the corresponding theoretical predictions.  The most important physical observation is a significant amplification of the velocity and vorticity fluctuations in the  critical and subcritical regimes (for $0.7  \leq \Omega \leq 0.9 $) with respect to the level typical for classical hydrodynamic   turbulence.  We consider this result as a manifestation of the \emph{enhancement of intermittency in  superfluid $^3$He turbulence}.

\subsubsection{2$\sp{nd}$-order structure functions of the velocity and vorticity  $S_2(r)$ and $T_2(r)$}

 Consider scaling behavior of the velocity 2$\sp{nd}$-order structure function $S_2(\tilde r)$ for different $\O$, shown in \Fig{f:3}a   as a function of a dimensionless distance $\tilde r=r/\eta$. For the  classical hydrodynamic turbulence ($\O=0${, black line),  $S_2(\tilde r)$ demonstrates the expected behavior: a viscous regime, with $S_2(\tilde r)\propto \tilde r^2$ for small $r$ followed by the K41 regime, with $S_2(\tilde r)\propto \tilde r^{\zeta_2} $, with $\zeta_2=2/3$ both shown by black dashed lines. Note, that intermittency correction to the K41 value of the scaling exponent $\zeta_2$ ($\zeta_2\approx 0.70$ instead of $\zeta_2=2/3\approx 0.67$) is not visible  on the scale of \Fig{f:3}a and will be discussed below. The spectrum for $\O=0.25$ (brown line) behaves similarly to the classical case $\Omega=0$, just with larger cross-over value of $\tilde r$. For larger subcritical values of $\O=0.5$ (red line) and $\O=0.7$ (green line), the viscous $S_2(\tilde r)\propto \tilde r^2$ behavior for small $r$ is now followed by an apparent scaling behavior $S_2(\tilde r)\propto \tilde r^{\zeta_2}$ with $\frac 23< \zeta_2<2 $.This is a consequence of apparent scaling behavior of the subcritical LNV spectrum\,\eqref{LNV-a}, discussed in the Introduction.   For example,for $\Omega=0.5$  $\zeta_2\approx 1.0$, while for $\Omega=0.7$, the apparent exponent  $\zeta_2\approx 1.4$, and become close to $\zeta_2\approx 2$ already for the near critical value of  $\Omega=0.9$. Note that for much larger Reynolds numbers, these apparent exponents are expected to appear only around $r_\times \sim 1/k_\times$.  For $r\ll r_\times$ the apparent exponent should  approach the classical value $\zeta_2=2/3$ and for $r\gg r_\times$ -- the critical value $\zeta_2=1$.

As explained in \Sec{sss:struct}, in the supercritical regime, when  $ E\sb s(k)\propto k^{-x}$ with $x>3$, the integral\,\eqref{9} losses its locality and is dominated by small $r$, where the velocity field can be considered as smooth. In this regime the viscous behavior $S_2(\tilde r)\propto \tilde r^2$ is expected for all $\Omega \geqslant 0.9$, as is confirmed in \Fig{f:3}a.

Moreover, in this case the scaling behavior of the velocity structure function $S_2(\tilde r)\propto \tilde r^2$ is disconnected from the energy scaling $ E\propto k^{-x}$. The vorticity structure function $T_2(\tilde r)$ is more informative for this regime, because, as shown in \Sec{sss:struct}, the vorticity field is not smooth for $x<5$.

\Fig{f:3}b compares the behavior of $T_2(\tilde r)$ for different $\O$. Consider first the test case $\O=0$, shown by a black line.  For very small $\tilde r$, when $1/\tilde r$ exceeds viscous cutoff of the energy spectrum, we see the viscous behavior $\propto \tilde r^2$, followed by the saturation region $T_2(\tilde r)\simeq $const. As explained in \Sec{sss:struct}, this is because the energy spectrum exponent $x=5/3$ is\emph{ below} the  lower edge of the  vorticity locality window\,\eqref{T2a}. For $x<3$, the integral\,\eqref{8b2} is dominated by large $k$ in the interval $\dfrac \pi r < k < k\sb {max}$ and $T_2(\tilde r)$ becomes $r$-independent, as observed.

In \Figs{f:3} we present two cases with $x$ within the locality window for vorticity\,\eqref{T2a}, $3< x < 5$: $\O=0.9$ with $x\approx 3$ and $\O=1.1$ with $x\approx 3.66$. According to our asymptotical (for infinitely large scaling interval) prediction\,\eqref{T2b}, we expect for these cases $z\approx 0$ and $z\approx 0.66$. The numerically found values (see \Fig{f:3}b) are slightly larger:  $z\approx 0.3$ and $z\approx 0.8$. Having  relatively short scaling interval, we consider this agreement as acceptable.

For even stronger mutual friction $\O=2.5$ and $\O=5$, the energy scaling exponent $x\approx 6.9$ and $x\approx 8.6$, are \emph{above} the upper edge of the vorticity locality window\,\eqref{T2a}. In this case integral\,\eqref{8b2} diverges at lower limit, giving
\begin{equation}\label{T2low}
T_2(\tilde r)\simeq \frac 4 3 \, \tilde r^2 \int\limits _{k\sb {min}}^\infty  k^4 E(k)\, dk \propto \tilde r^2\ .
\ee
as is indeed observed in \Fig{f:3}b.

\begin{figure*}
\begin{tabular}{ccc}
(a) & (b) & (c) \\
\includegraphics[scale=0.28]{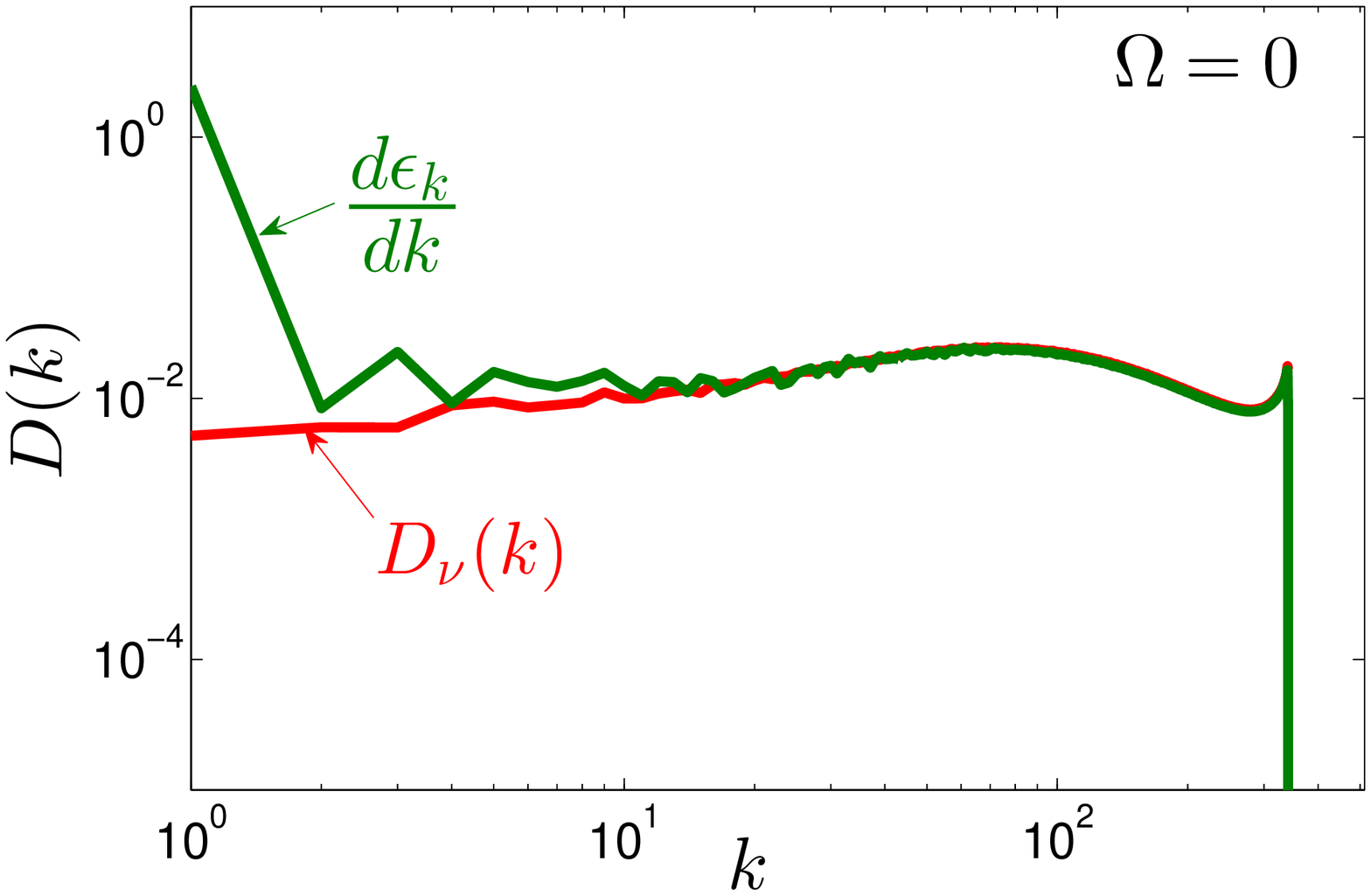}&
\includegraphics[scale=0.28]{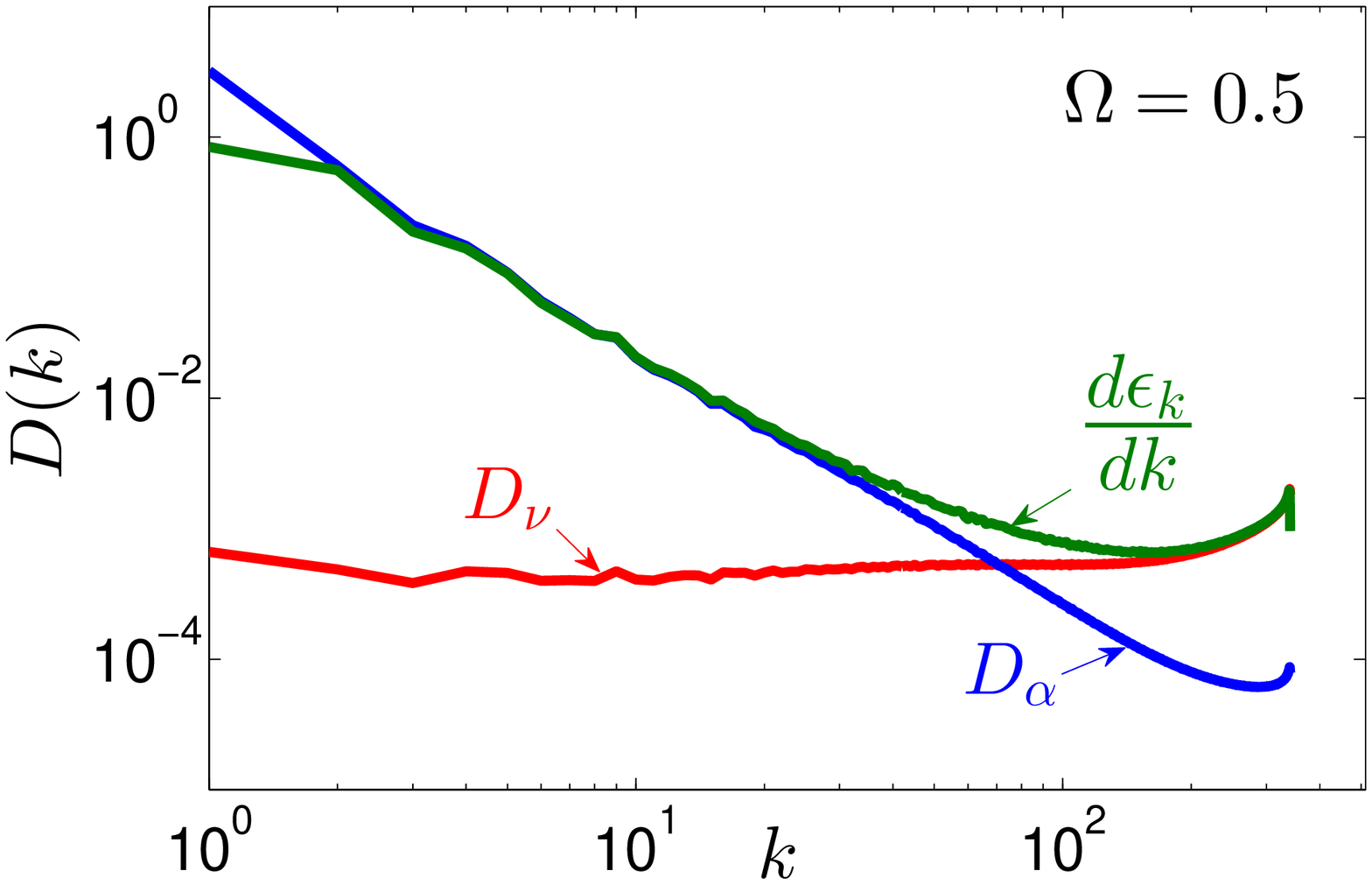}&
 \includegraphics[scale=0.28]{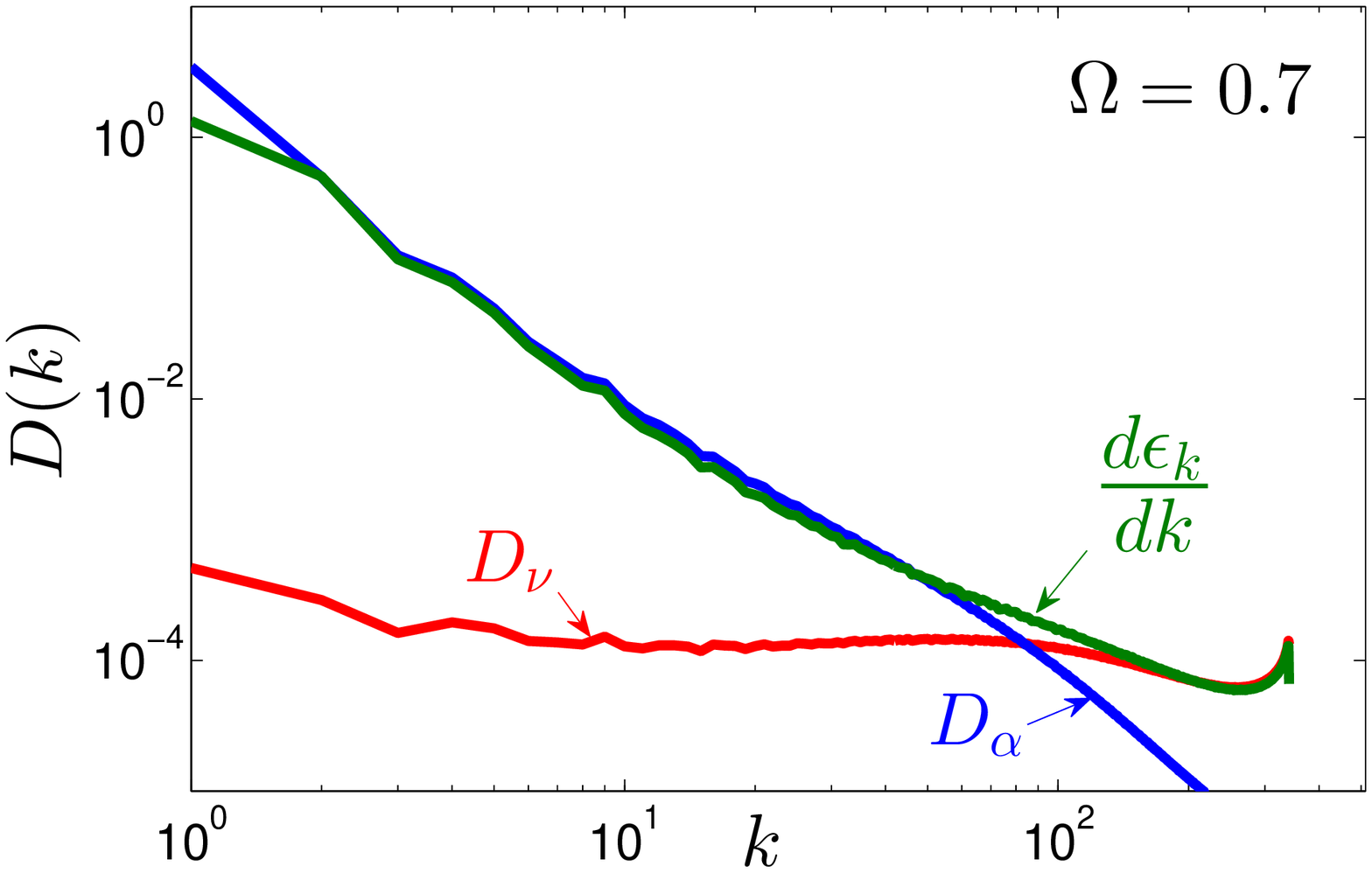}\\
 (d) & (e) & (f) \\
\includegraphics[scale=0.28]{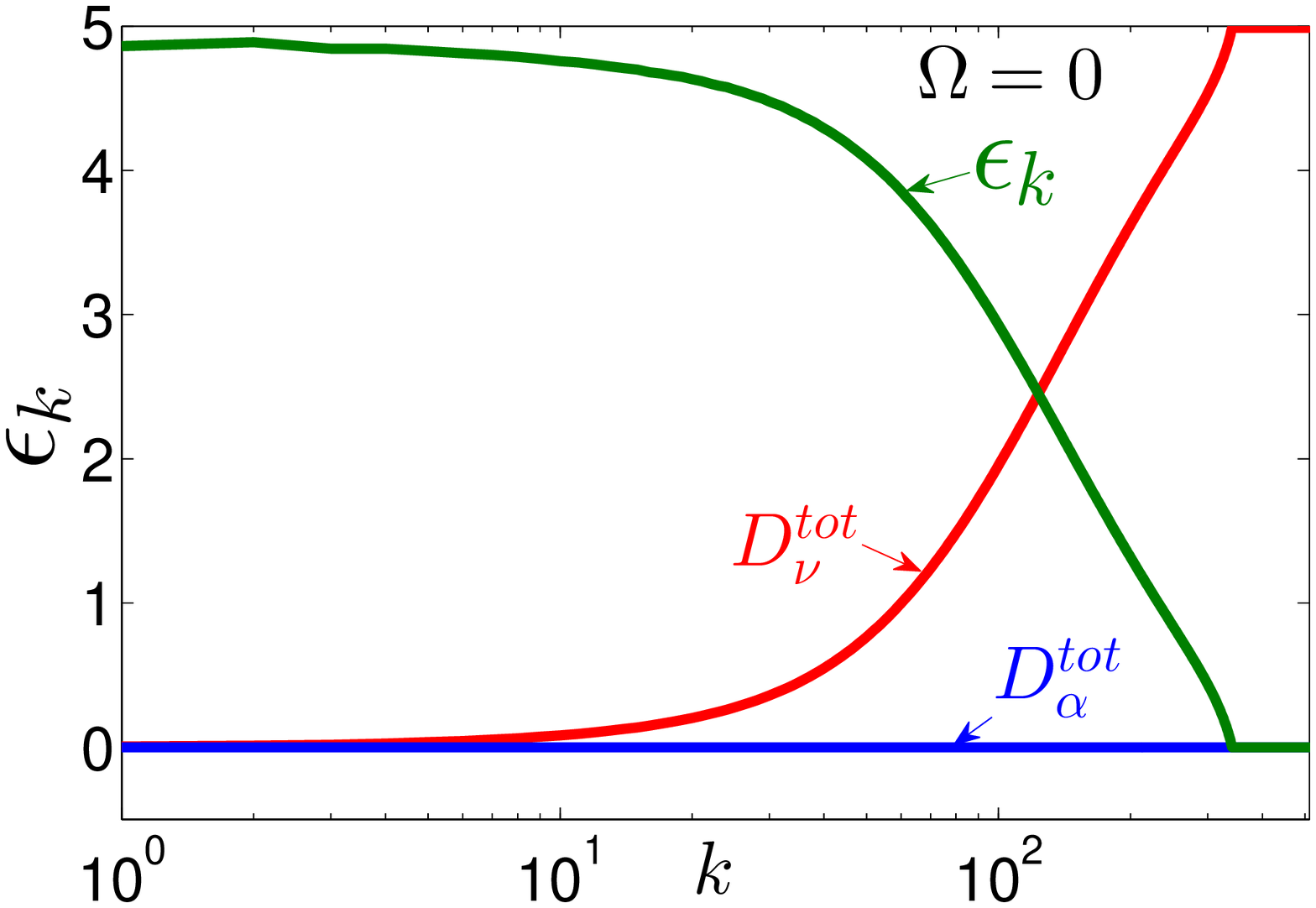}&
\includegraphics[scale=0.28]{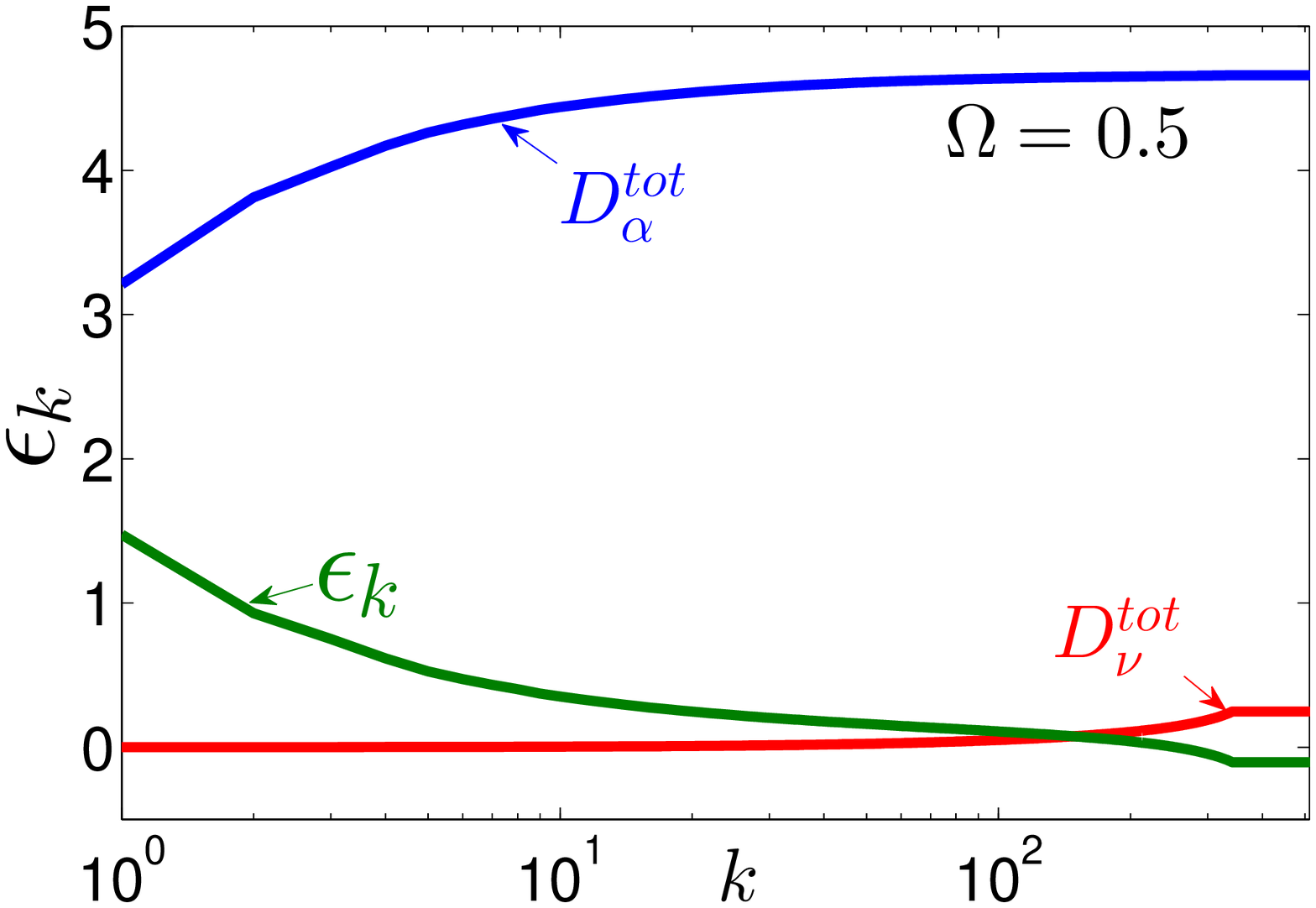}&
 \includegraphics[scale=0.28]{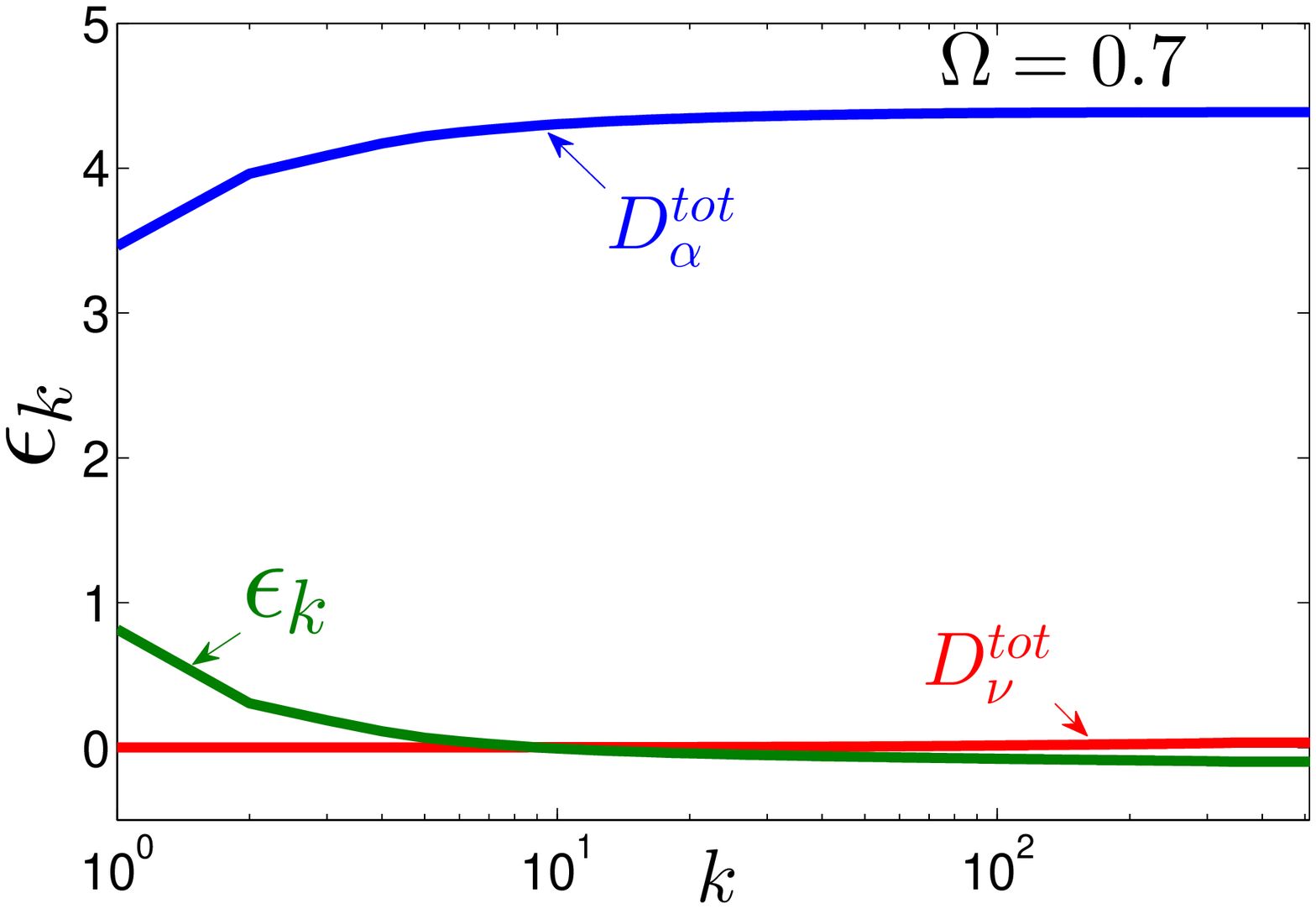}\\

               \end{tabular}
\caption{\label{f:5} Color online. The differential [Panels (a),(b),(c)] and the integral [Panels (d),(e),(f)] energy balances in the subcritical regimes with $\Omega =0$ [Panels (a),(d)], $\Omega =0.5$ [Panels (b),(e)] and  $\Omega =0.7$ [Panels (c),(f)]. The nonlinear energy transfer is shown by green lines, the viscous dissipation by red lines and the dissipation by mutual friction by blue lines.}
\end{figure*}

 \subsubsection{\label{sss:enhan}4$\sp{nd}$-order structure functions,  flatnesses and enhancement of intermittency}
 Consider now 4$\sp{th}$-order structure functions of the velocity and vorticity  $S_4(\tilde r)$ and $T_4(\tilde r)$, shown in \Fig{f:3}c and \Fig{f:3}d, for the  subcritical and supercritical regimes. As is well known, for the Gaussian statistics  or, in a more general case, for  the ``mono-scaling"  statistics,  the fourth-order structure functions are proportional to the square of the second one: $S_4(\tilde r)\propto S_2^2(\tilde r)$ and $T_4(\tilde r)\propto T_2^2(\tilde r)$.
 We find such a behavior  for very small $\tilde r$.  For the classical case  $\Omega=0$ (\Fig{f:3}c)  we see again  scaling   exponent  $\zeta_4$ close to the standard K41 value $4/3\approx 1.33$  with intermittency corrections, hardy visible  on this scale. For larger $\O$,  the subcritical LNV spectrum\,\eqref{LNV-a} becomes a superposition of two scaling laws  and, as we mentioned in the Introduction, in the vicinity of a crossover wave number $k_\times$ may be  approximated as $k^{-x}$ with an apparent scaling exponent $\frac53 < x(k) < 3$. Indeed, we see   in \Fig{f:3}c that  the apparent value of $\zeta_4$ definitely deviate from 4/3,  approaching,  for example, $\zeta_4\approx 1.7$ for $\Omega=0.5$ and  $\zeta_4\approx 2.3$ for $\Omega =0.7$.  Such a steepening of the structure functions spectra is caused by the energy dissipation by mutual friction (see \Fig{f:1}).

 More importantly, upon increase in $\Omega$ the   apparent  scaling of the velocity field progressively deviates  from the self-similar behavior  type with $S_4(\tilde r)\propto S_2^2(\tilde r)$ and $\zeta_4= 2\zeta _2$.  For example, for $\Omega=0.5$    $\zeta_4\approx 1.7 <  2 \zeta_2 \approx 2.0$ (such that $ \xi=2\zeta_2 - \zeta_4\approx 0.3$) and  for $\Omega=0.7$ the difference    $\xi \approx 0.5$.

 To  further detail  this multiscaling regime, we  plot in \Figs{f:4} the velocity and vorticity flatnesses  $F_v(\tilde r)$ and  $F_\omega(\tilde r)$, defined as:
  \begin{equation}\label{flat}
   F\sb v(\tilde r)= S_4(\tilde r)/ S_2^2(\tilde r)\,, \quad  F_\o(\tilde r)= T_4(\tilde r)/ T_2^2(\tilde r) \ .
   \end{equation}
For the Gaussian and mono-scaling statistics, $F\sb v(\tilde r)$ and $F_\omega(\tilde r)$ must be $\tilde r$-independent.
In particular, for the Gaussian statistics $F\sb v(\tilde r)=F_\omega(\tilde r)$=3.
As is evident in \Fig{f:4}a and \Fig{f:4}b, the intermittency corrections, hardly visible for structure functions for $\Omega=0$, are clearly exposed by the flatness.   The velocity flatness $F_v(\tilde r)$ for this case (black solid line in \Fig{f:4}a) approximately follow the   intermittent   exponent for turbulence in classical fluids  $\xi\sb{cl} \approx 0.15$, which is close to the experimental values for both the longitudinal and transversal structure functions (for previous experimental and numerical works on intermittency in the classical space-homogeneous isotropic turbulence see Refs. \cite{ Gotoh02,Chanal00,Kaha98,Benzi10,Ishihara07,BTS-97}) .  As the mutual friction become stronger, the apparent exponent $\xi$   increases, reaching its maximum $\xi\sb{max} \approx 0.45\approx 3\xi\sb{cl}$  at  $\Omega=0.7$.  The vorticity flatness $F_\omega(\tilde r)$[\Fig{f:4}b] too reaches its maximum for small $\tilde r$ at slightly larger value of  $\Omega\approx 0.9$. This is a clear evidence of significant enhancement of intermittency in the near-critical regimes of superfluid $^3$He turbulence.

 Additional important information can be found in \Figs{f:4}c and \ref{f:4}d, where $\Omega$-dependence of the velocity and vorticity flatnesses is shown for different $\tilde r$. The sharp peak appears for $\Omega\lesssim 0.9$.  In the small $\tilde r$ range, the velocity flatness $F\sb v(\tilde r)$ for $\Omega=0.7$ reaches value about 25 (compare with the Gaussian value of three and the classical hydrodynamic value about seven). At the same time the vorticity flatness  reaches value of about 200,  exceeding the Gaussian limit by almost two orders of magnitude.

 In the supercritical regime, the intermittency sharply decreases. For example,  for $\Omega>2.5$ the velocity flatness drops even below the Gaussian limit, indicating that the time dependence of the velocity becomes  sub-Gaussian.

\subsection{\label{ss:balance} Energy balance}
The direct information about the relative importance of the energy dissipation by the effective viscosity and by the mutual friction can be obtained from an analysis of the energy balance, shown in \Figs{f:5}. The energy balance for the classical turbulence ($\Omega=0$) is presented in \Fig{f:5}a. As expected, the energy input at a shell with a given wave number $k$,   Tr$(k)=  d \ve(k)/dk$ (green line) is compensated by the viscous dissipation D$_\nu=2\nu\sb s  E\sb s(k)$ (red line). The discrepancy in the region of very small $k$ is caused by the energy pumping, which is not accounted in the balance \Eq{BALa}. Sometimes it is more convenient to discuss a ``global" energy balance, analyzing instead of the ``local" in $k$ balance \Eq{BALa} its integral from $k=$ to a given $k$. In the stationary case this gives:
\begin{subequations}\label{bal-g}
\begin{eqnarray}\label{bal-gA}
\ve(k)&=&\ve_0- \mbox{D}_\nu\sp{tot}(k) - \mbox{D}_\alpha\sp{tot}(k)\,,
 \\ \label{bal-gB}
\mbox{D}_\nu\sp{tot}(k)&=&\int_0^k \mbox{D}_\nu (q)  dq, \ \mbox{D}_\alpha\sp{tot}(k)=\int_0^k \mbox{D}_\alpha (q)  dq\,. ~~~~~~
\end{eqnarray}
\end{subequations}
As we see in \Fig{f:5}d (for $\Omega=0$), the energy flux over scales $\ve(k)$ is almost constant up to $k\simeq 20$ and then decreases due to the viscous dissipation. Accordingly, $E\sb s(k,0)$, shown in \Fig{f:1}a by black solid line, exhibits a  K41 scaling $\propto k^{-5/3}$. Minor upward deviation  from this behavior may be a numerical artifact.

The energy balance in the subcritical regime of the superfluid $^3$He turbulence, shown in \Figs{f:5}b and \ref{f:5}e for $\Omega=0.5$ and  in \Figs{f:5}c and \ref{f:5}f for $\Omega=0.7$  demonstrates a qualitatively different behavior. We see in \Figs{f:5}b and \ref{f:5}c that for almost all wavenumbers, the energy input Tr$(k)$ in a given $k$ (shown by green lines) is balanced by the mutual friction dissipation D$_\alpha(k)$ (shown by the blue lines). Only for large $k \gtrsim 75$, the viscous dissipation begin to dominate. Nevertheless, as seen in \Figs{f:5}e and \ref{f:5}f, the total contribution to the energy dissipation is dominated by the mutual friction everywhere.
As expected, for larger and larger $\Omega$ the crossover wave number  $k_\times$, at which the local dissipation by viscosity and by mutual friction are equal, increases (compare \Fig{f:5}b with $\Omega=0.5$ and  \Fig{f:5}c with $\Omega=0.7$) and   reaches $k\sb{max}$ for the critical regime with $\O=0.9$(\Fig{f:6}a). In this case the viscous and the mutual friction dissipation become compatible only for $k\simeq k\sb{max}$.

In the supercritical regime, shown in \Figs{f:6}e and \ref{f:6}f for  $\Omega=1.1$ and $\Omega=5$, the contribution of the viscous dissipation (red lines) becomes less and less important with the increase in the supercriticality.   In these cases, the nonlinear input to  the   energy,  Tr$(k)=d \ve(k)/dk$ (green lines) is fully compensated by the mutual friction dissipation(blue lines).

The global energy balance, shown in \Figs{f:6},  confirms this physical picture.

\begin{figure*}
\begin{tabular}{ccc}
(a) & (b) & (c) \\
\includegraphics[scale=0.28]{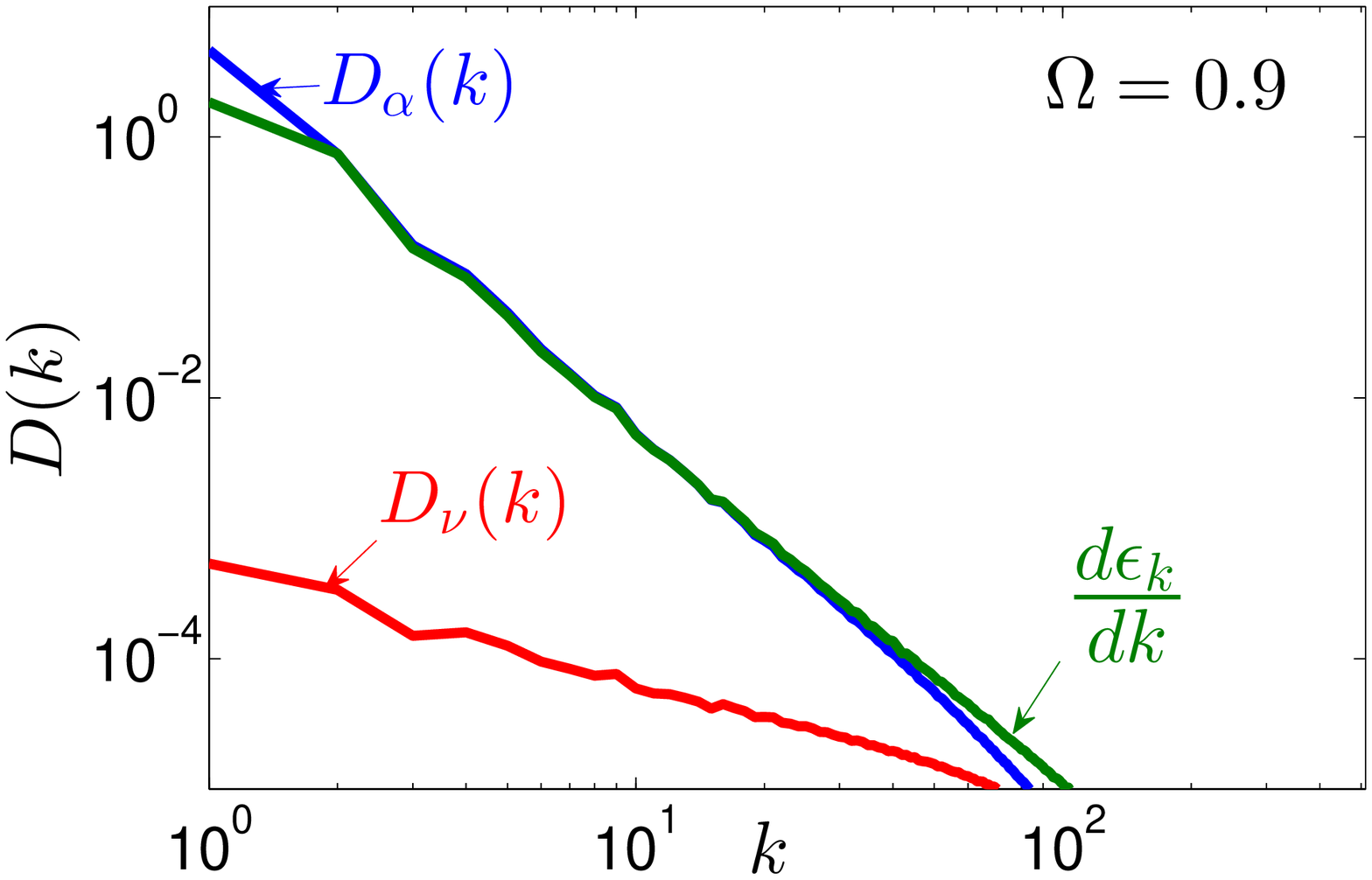}&
\includegraphics[scale=0.28]{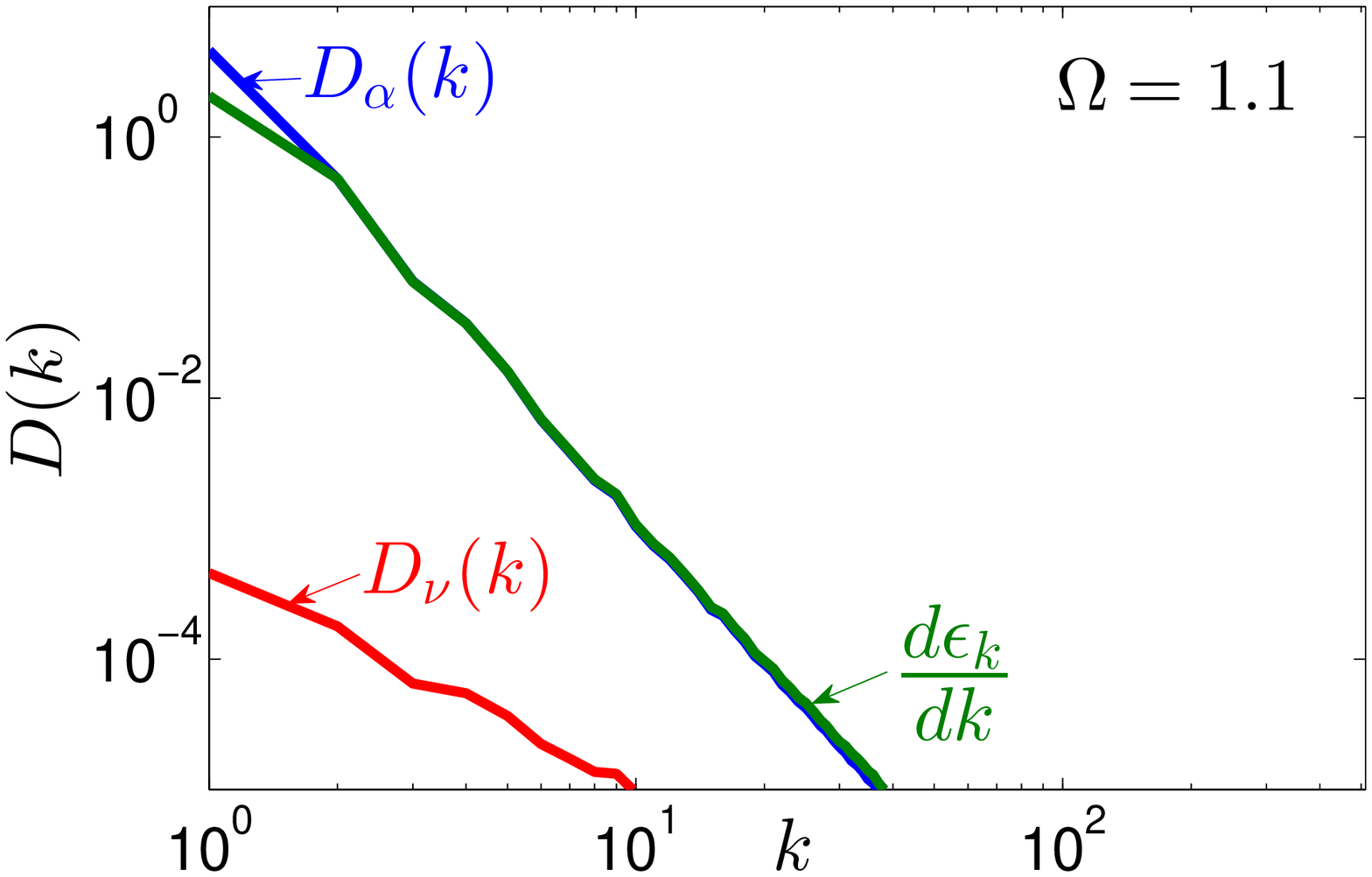}&
 \includegraphics[scale=0.28]{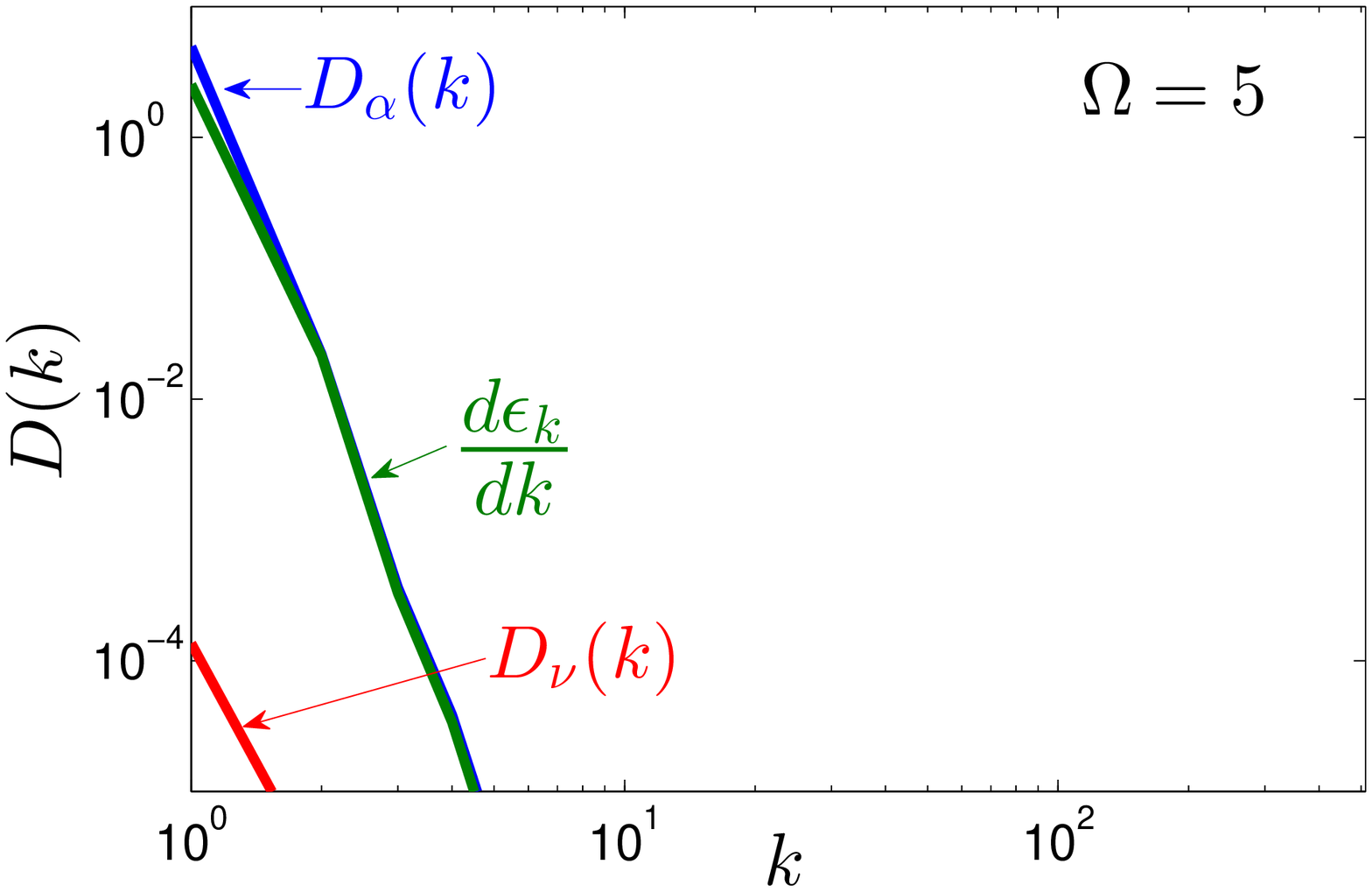}\\

 (d) & (e) & (f) \\
 \includegraphics[scale=0.28]{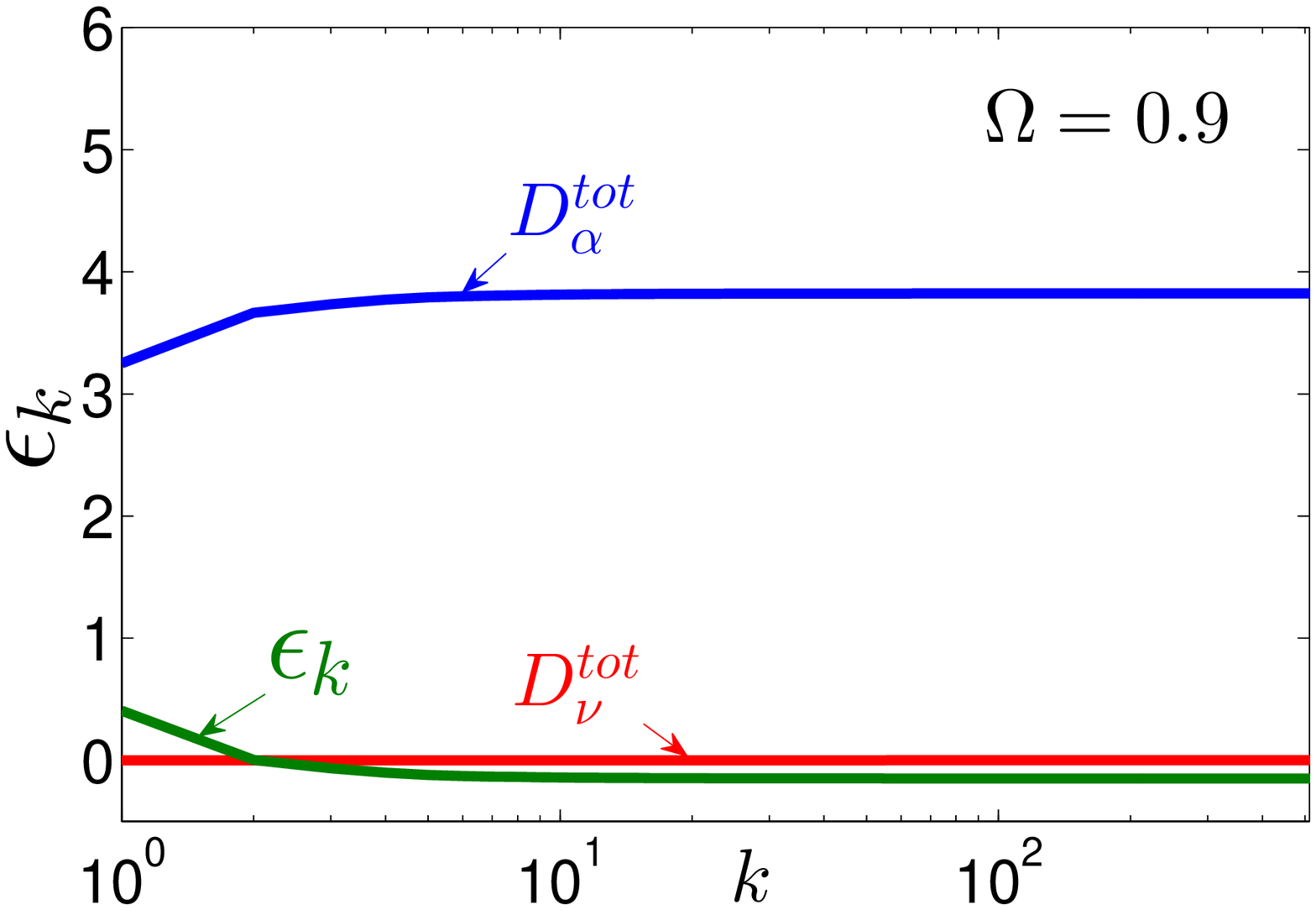}&
\includegraphics[scale=0.28]{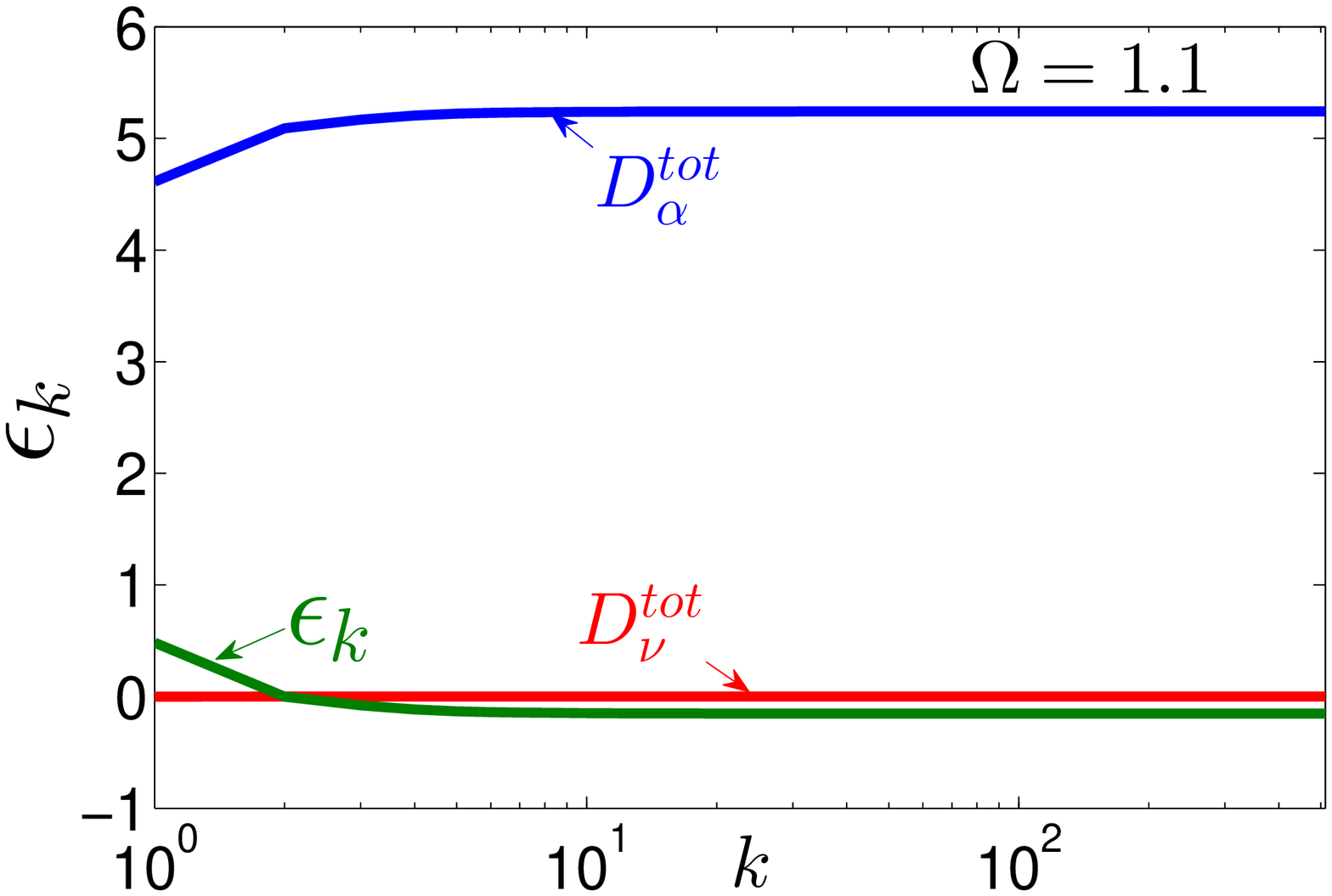}&
 \includegraphics[scale=0.28]{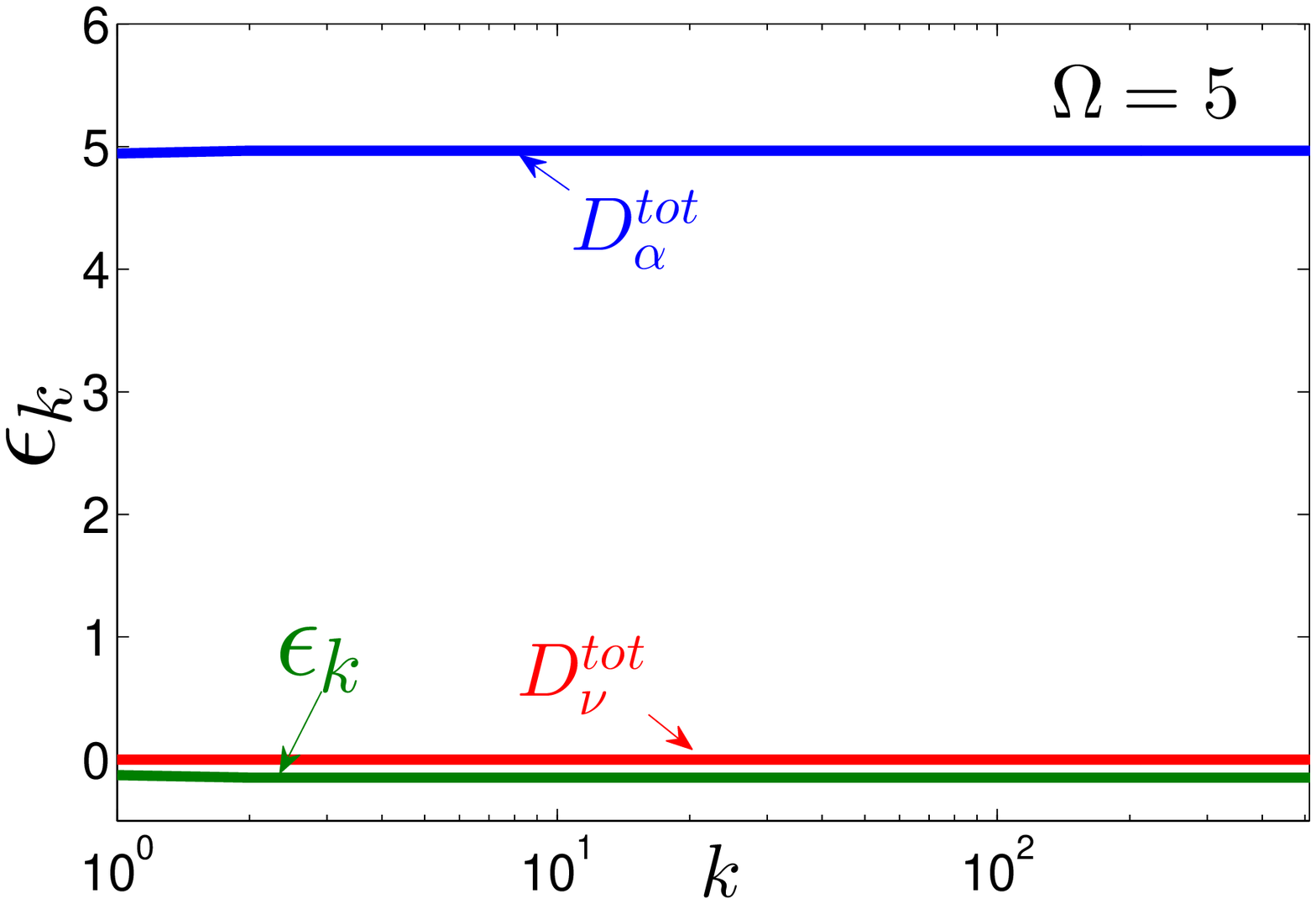}\\
 \end{tabular}
\caption{\label{f:6} Color online. The differential [Panels (a),(b),(c)] and the integral [Panels (d),(e),(f)] the energy balances in the critical and the supercritical regimes with $\Omega =0.9$ [Panels (a),(d)], $\Omega =1.1$ [Panels (b),(e)] and $\Omega =5.0$ [Panels (c),(f)]. The nonlinear energy transfer is shown by green lines, the viscous dissipation by red lines and the dissipation by mutual friction by blue lines.}
\end{figure*}

\begin{figure*}
\begin{tabular}{ccc}
(a) & (b) & (c) \\
\includegraphics[scale=0.27]{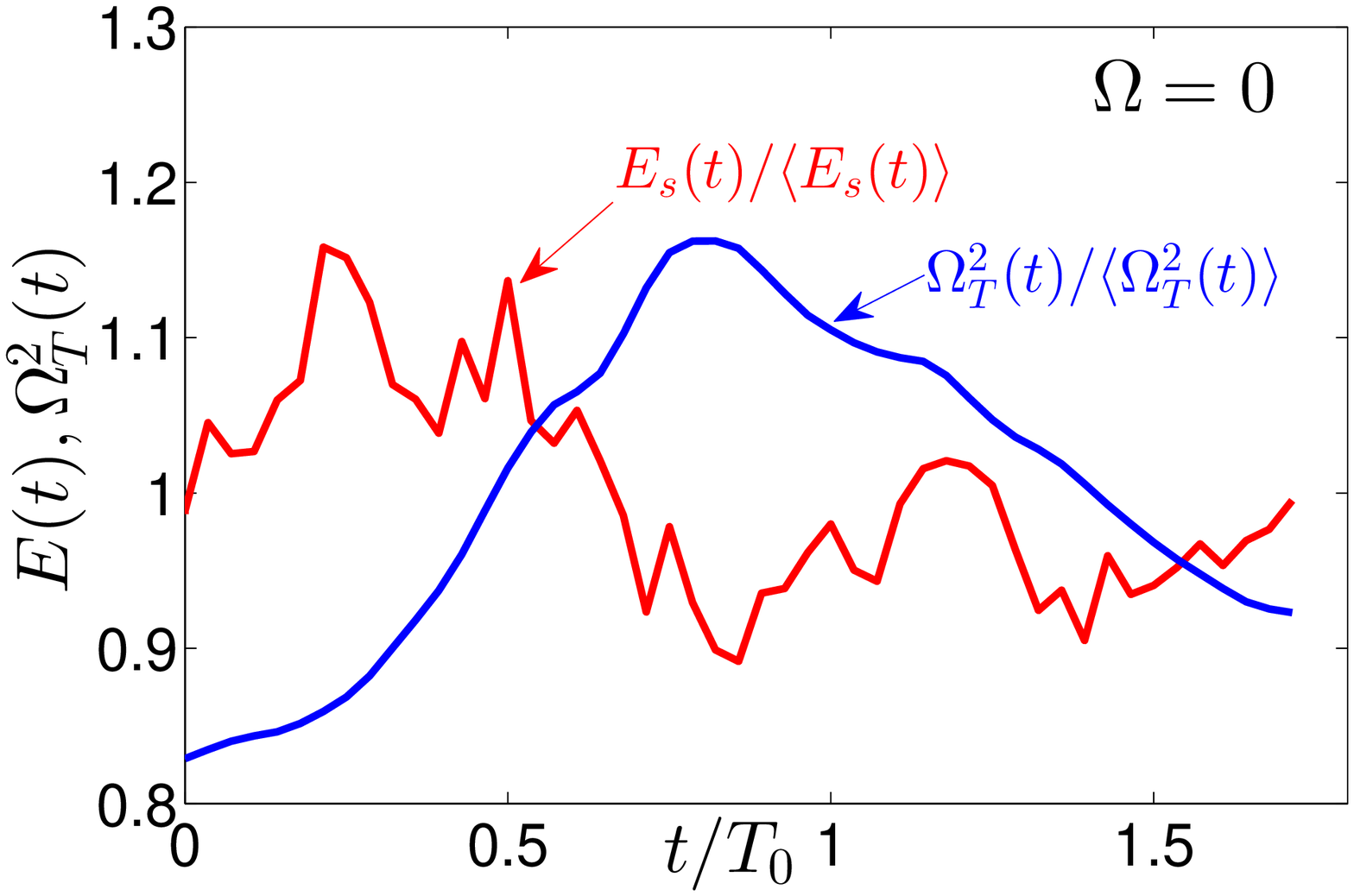}&
 \includegraphics[scale=0.27]{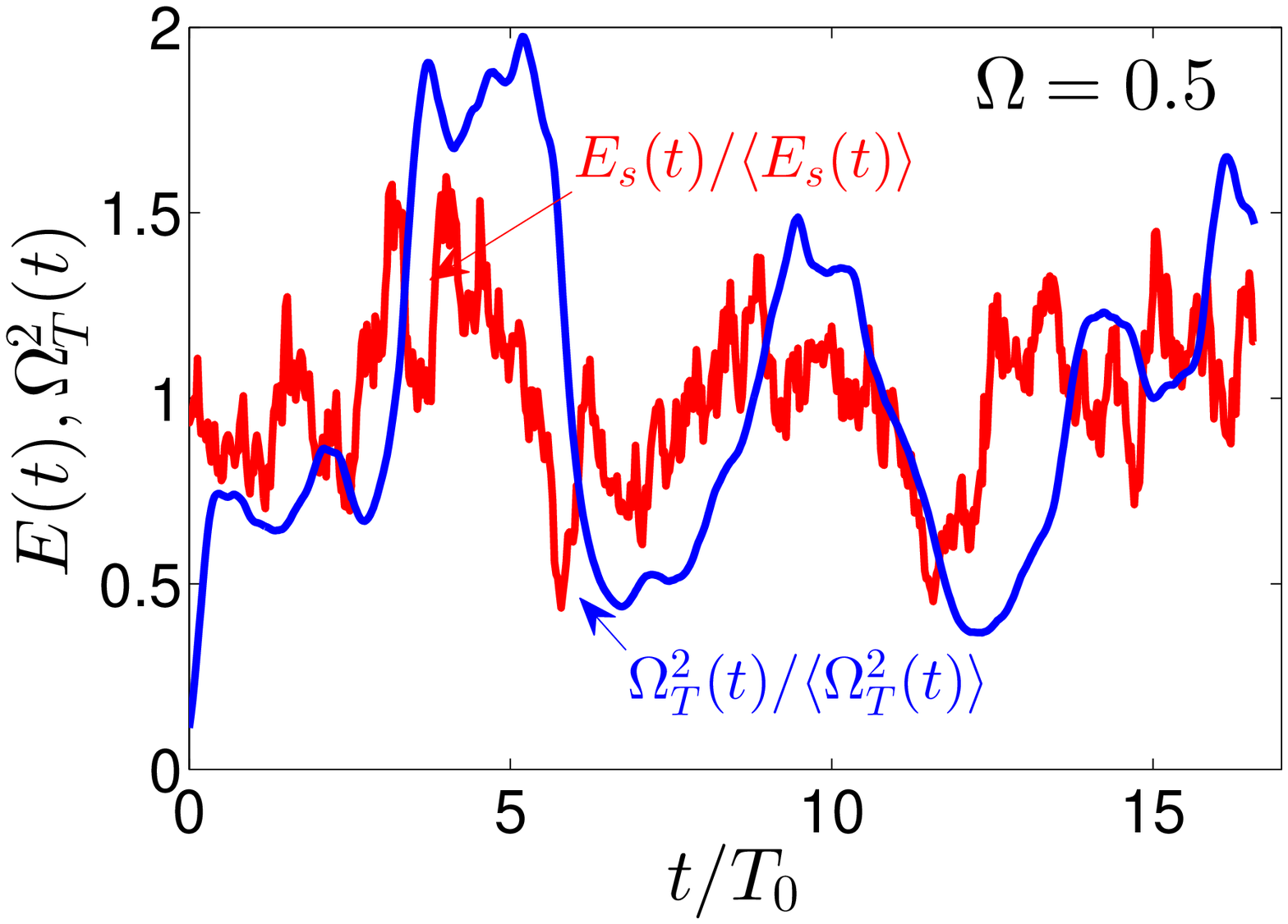}&
\includegraphics[scale=0.27]{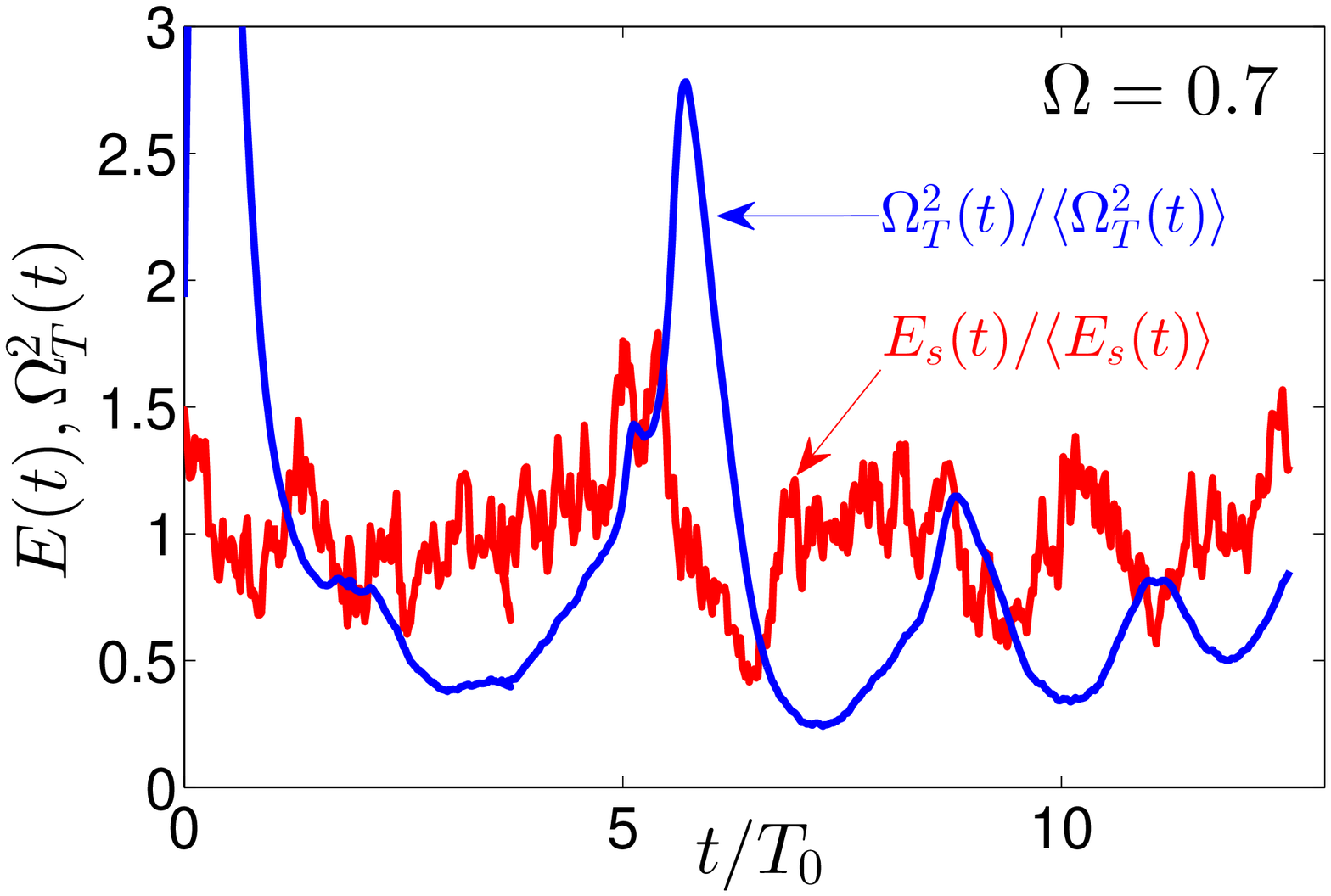}\\
(d) & (e) & (f) \\
\includegraphics[scale=0.27]{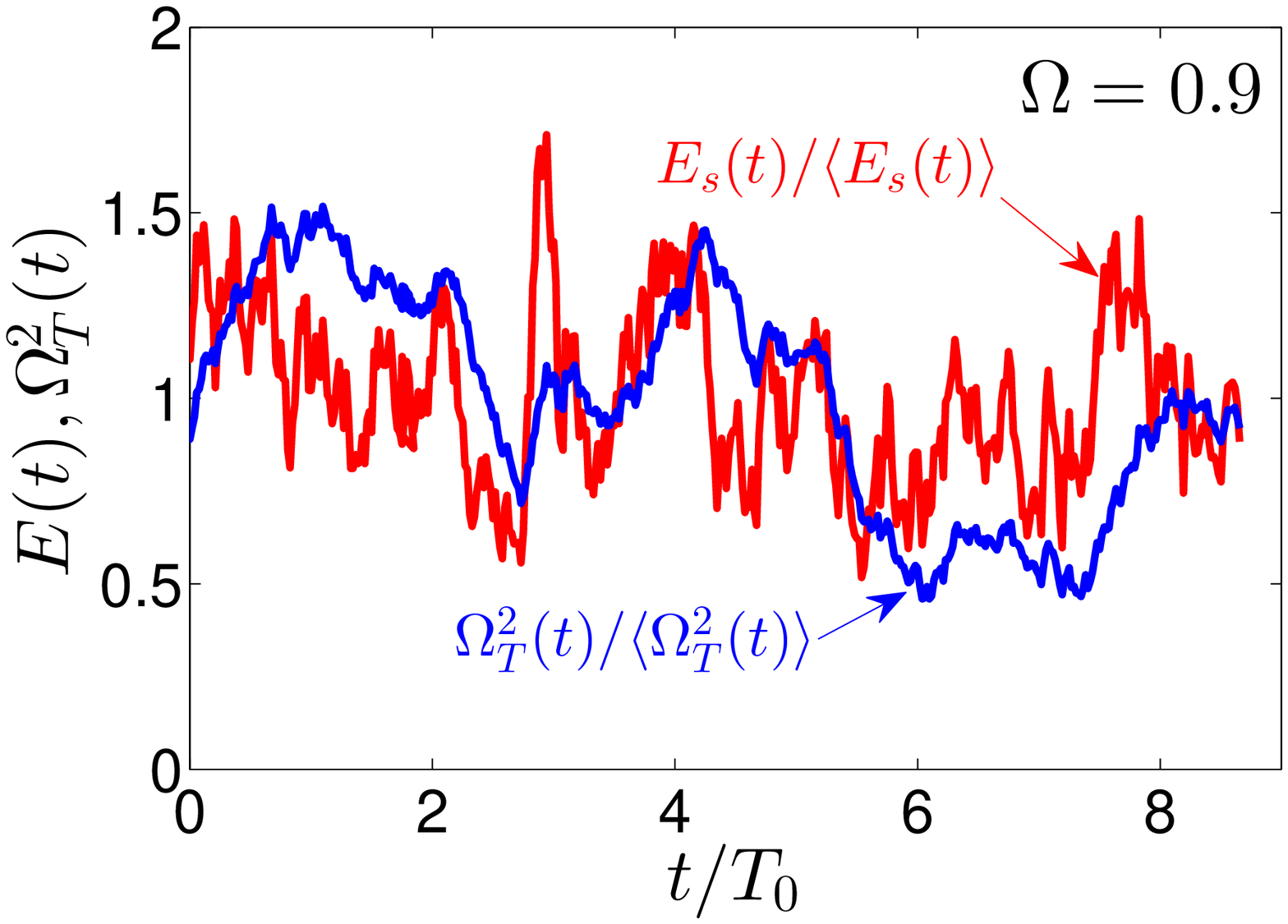}&
\includegraphics[scale=0.27]{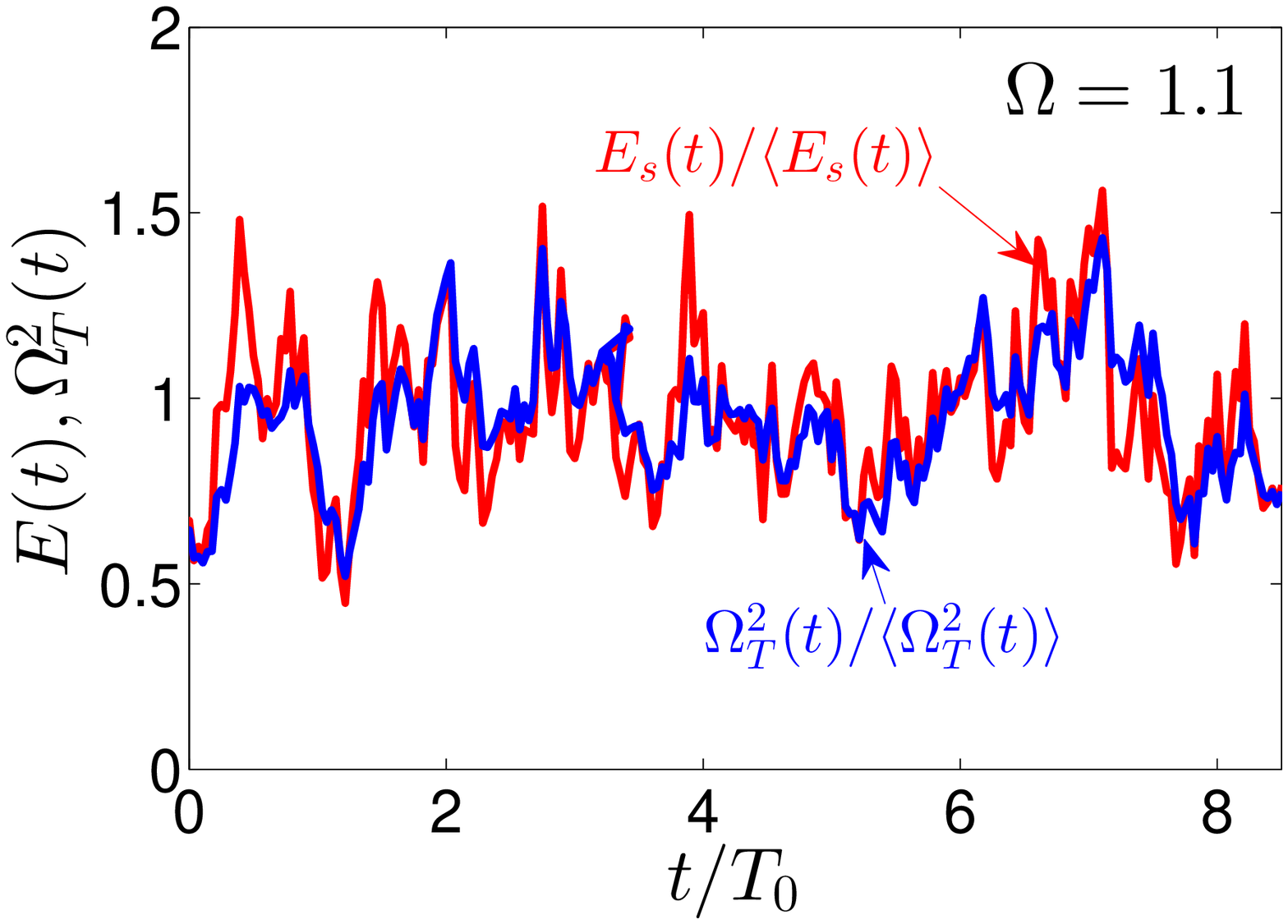}&
\includegraphics[scale=0.27]{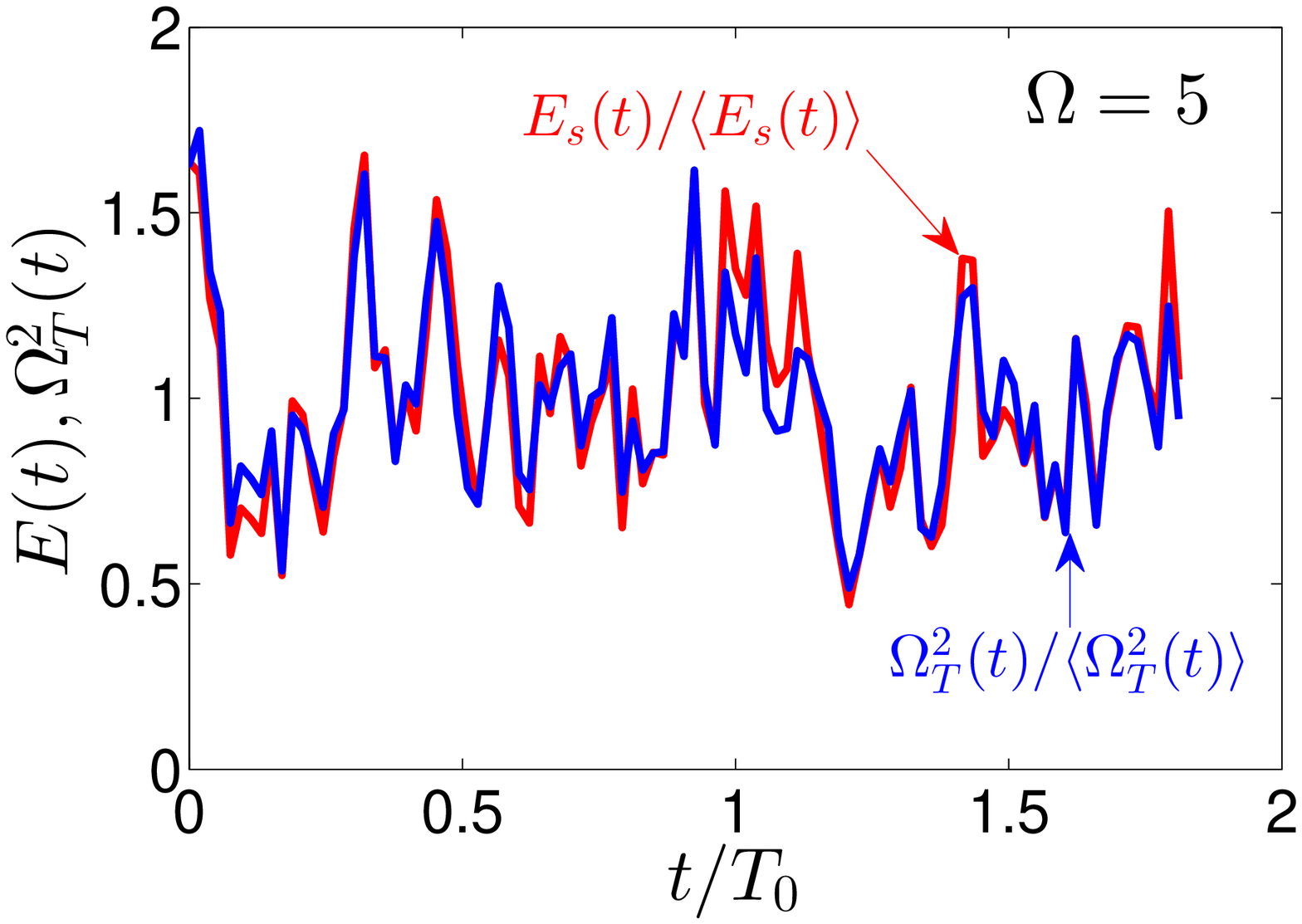}\\
\end{tabular}
\caption{\label{f:7} Color online. The energy (red lines) and enstrophy (blue lines) time evolutions in the subcritical regime normalized by mean-in-time values [Panels (a),(b),(c) with \ $\Omega=0, \ 0.5, \ 0.7$], the critical [Panel (d), $\Omega=0.9$] and the supercritical regime [Panels (e),(f) with $\Omega=1.1,\ 5.0$].}
\end{figure*}

\subsection{\label{ss:evolution} Energy and enstropy time evolution}
We consider here  evolution of the total superfluid energy $E\sb s(t)$ 
\begin{subequations}\label{comp}
  \begin{eqnarray}\label{compA}
  E\sb s(t)&=&\int E\sb s(k,t) dk
\end{eqnarray} \end{subequations}
and enstrophy $1/2\Omega^2 \Sb T (t)$.
As expected, in  the subcritical regime,  when $  E(k,t)$ has apparent slope $\propto k^{-x}$ with  $\frac 53 < x(k) < 3$, the integral\,\eqref{compA} for total energy $E(t)$ is dominated by the small $k\sim k\sb{min}$, while the integral\,\eqref{ot} for total enstropy $\Omega^2\Sb T(t)$ is dominated by the large $k\sim k\sb{max}$. Therefore, for the large ratio $k\sb{max}/k\sb{min}$ (in our case  $k\sb{max}/k\sb{min}\sim 10^3$), one expects an uncorrelated behavior of $E(t)$ and $\Omega^2\Sb T(t)$ in case of a well developed turbulent cascade. This behavior is confirmed in \Figs{f:7}a, \ref{f:7}b and \ref{f:7}c.   In the supercritical regime, with the slope $x>3$, both $E(t)$ and $\Omega^2\Sb T(t)$ are dominated by the small $k\sim k\sb{min}$ and have to be well correlated, as is indeed seen in \Figs{f:7}e and \ref{f:7}f.   However, in the critical regime (\Fig{f:7}d   with $\O=0.9$), $E(t)$ and $\Omega^2\Sb T(t)$ are  still uncorrelated because $E(t)$  is dominated by $k\sim k\sb{min}$, while $\Omega^2\Sb T(t)$ has equal contributions from all $k$.

\subsection{\label{ss:temp} Relation between $\O$ and temperature $T$ of possible experiments}
Up to now we have considered $\O$ as a free parameter that determines the mutual friction by  \Eq{1e}, in which $\Omega\Sb T$ is given by \Eq{ot}. After the simulation with a prescribed $\O$ was completed, we numerically computed $\Omega\Sb T^2$, using found energy spectra and \Eq{ot}, see Tab.\, \ref{t:1}. Now, using \Eq{1e} we can find $\~\alpha= \Omega/\Omega\Sb T$ for a given $\Omega$ in the simulations. The parameter $\~\a$ in \He3 strongly depend on temperature, as reported in \cite{Bevan} and shown in \Fig{f:8}.  Using these data, we can find $T$ corresponding to the simulations with any prescribed $\O$.

\begin{figure}[t]
\includegraphics[scale=1.1]{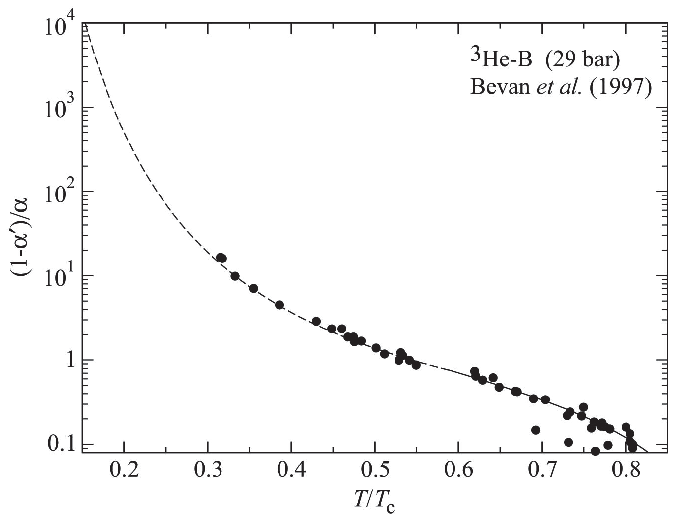}
\caption{\label{f:8} Temperature dependence of the mutual friction parameter $\~ \alpha(T)=\a/(1-\alpha')$, taken from \Ref{Bevan}}
\end{figure}

\section{\label{s:sum}Summary}
This paper examined the basic statistical properties  of the large-scale, homogeneous, steady, isotropic quantum turbulence in superfluid $^3$He,  developing further some previous results\,\cite{LNV,He3a}.
Direct numerical simulations of the gradually damped version of the HVBK  coarse-grained two-fluid
model of the superfluid  He, \Eqs{NSE}\cite{HV,BK,PRB} were performed using pseudo-spectral methods in a fully
periodic box with a grid resolution of $N = 1024^3$.  The analytic study was based on the LNR integral closure for the energy flux\,\cite{LNR}, \Eq{24}, adapted for $^3$He turbulence in \Eq{25}.    Both the DNS and the analytic approaches do not use the assumption of locality of the energy transfer between scales.
The main findings are:

1. The direct numerical simulations confirmed the previously found\,\cite{LNV,He3a}  subcritical\,\eqref{LNV-a} and  critical\,\eqref{LNV-cr} energy spectra and showed that for   $T < 0.37 \, T\sb c$  (see Tabl.\,\ref{t:1})  the analytic prediction are in a good quantitative agreement with the DNS results, using a single fitting parameter $b$ for all temperatures.   The reason for this agreement is that in the subcritical regime the energy transfer over scales is indeed local, in accordance with the  basic assumptions in \Refs{LNV, He3a}. In the critical regime\cite{LNV,He3a} with $E(k)\propto k^{-3}$,  the exact locality of the energy transfer fails: all the scales contribute equally to the transfer of energy to the turbulent fluctuations with a given $k$. This leads to a logarithmic corrections to the spectrum  $E(k)\propto k^{-3}$ that cannot be detected with our DNS resolution.

2.~For  $T>0.37 \, T\sb c$, when the mutual friction exceeds some critical value, we
observed in DNS and confirmed analytically the scale-invariant spectrum $E(k)\propto k^{-x}$ with a ($k$-independent) exponent $x > 3$. The exponent $x$  increases gradually with the temperature, reaching  in our simulation  the value $x\approx 9$  for $T\approx 0.72\, T\sb c$.    The reason for this behavior of the supercritical spectra with $x>3$ is that the energy is transferred directly to any given $k$ from the energy containing region at small $k$.

3. We analyzed the 2$\sp{nd}$-order structure functions of the velocity and
vorticity $S_2(r)$  and $T_2(r)$ and demonstrated   that  although  their $r$-dependence can be rigorously found from the energy spectrum $E(k)$, their $r$-dependence is much less informative that the $k$-dependence of $E(k)$.

4. The 4$\sp{nd}$-order structure functions of the velocity and vorticity $S_4(r)$  and $T_4(r)$ provide important additional [with respect to $E(k)$] information about the statistics of quantum turbulence in the superfluid $^3$He. We discover a strong  enhancement of intermittency  in the near-critical regimes with  the level of turbulent fluctuations exceeding the corresponding level in the classical  turbulence by about an order of magnitude.

5. The analysis of the energy balance  and of  the energy and enstrophy time evolution in various (subcritical, critical and supercritical) regimes, confirms the discovered physical picture of the quantum $^3$He turbulence with the local and non-local energy transfer, in which the relative importance of the energy dissipation by the effective viscosity and by the mutual friction depends in a predicted  way on the temperature and the wavenumber.

We propose that these analytic and numerical findings in the description of
the statistical properties of steady, homogeneous, isotropic and incompressible turbulence of superfluid
$^3$He should  serve as a basis for further studies
of superfluid turbulence in more complicated or/and
realistic cases: anisotropic turbulence, transient regimes,
two-fluid turbulence of thermally driven counterflows in
superfluid $^4$He turbulence, etc.


\end{document}